\documentclass[twocolumn,journal]{IEEEtran}

\IEEEoverridecommandlockouts

\usepackage{cite}
\usepackage{amsmath,amssymb,amsfonts}
\usepackage{algorithmic}
\usepackage{graphicx}
\usepackage{textcomp}
\usepackage{xr}
\usepackage[table]{xcolor}
\usepackage[caption=false,font=footnotesize]{subfig}

\def\BibTeX{{\rm B\kern-.05em{\sc i\kern-.025em b}\kern-.08em
		T\kern-.1667em\lower.7ex\hbox{E}\kern-.125emX}}

\usepackage{tikz}
\usetikzlibrary{backgrounds}
\usetikzlibrary{spy}
\usetikzlibrary{calc,arrows.meta}
\usepackage{pgfplots}
\usetikzlibrary{plotmarks}
\usetikzlibrary{arrows.meta}
\usetikzlibrary{positioning,fit}
\usetikzlibrary{intersections}
\usetikzlibrary{patterns}
\usetikzlibrary{decorations.shapes}

\usepgfplotslibrary{patchplots}
\usepgfplotslibrary{colormaps}

\usepackage{cancel}

\usepackage{pdfpages} 

\usepackage{multirow}
\usepackage{makecell}
\usepackage{float}

\usepackage{algorithmic}
\usepackage[ruled,vlined]{algorithm2e}
\SetKwRepeat{Do}{do}{while}

\newcommand{\exportFigures}{true}
\newcommand{\exportFiguresAsPNG}{true}

\ifthenelse{\equal{\exportFigures}{true}}
{
	\usepgfplotslibrary{external}
	\tikzexternalize[prefix=compiled_tikz_figures/,optimize command away=\includepdf]
	\ifthenelse{\equal{\exportFiguresAsPNG}{false}}
	{
		\tikzset
		{   png export/.style={
				external/system call={
					pdflatex \tikzexternalcheckshellescape -halt-on-error --extra-mem-top=10000000 -interaction=batchmode -jobname "\image" "\texsource" && pdftops -eps "\image.pdf" && convert -density 700 -transparent white "\image.pdf" "\image.png"
		}}}
		\tikzset{png export}
	}
	{}
}
{}

\usepackage{url}

\usepackage{bm}
\usepackage[caption=false,font=footnotesize]{subfig}
\usepackage{caption}
\usepackage[normalem]{ulem}

\usepackage{acronym}

\usepackage{booktabs}
		
\usepackage{pdfpages}

\usepackage{tikz}
\usetikzlibrary{shapes,arrows}
\usetikzlibrary{arrows.meta}
\usetikzlibrary{positioning}
\usetikzlibrary{spy,backgrounds}
\usetikzlibrary{math}

\usepackage{ellipsis}

\usetikzlibrary{calc}
\usetikzlibrary{decorations.pathreplacing,decorations.markings,shapes.geometric}
\usetikzlibrary{decorations.pathmorphing}
\usetikzlibrary{fit}
\usetikzlibrary{pgfplots.groupplots}

\usetikzlibrary{calc,arrows.meta}

\usetikzlibrary{backgrounds} 

\definecolor{green(pigment)}{rgb}{0.0, 0.65, 0.31}
\definecolor{frenchblue}{rgb}{0.0, 0.45, 0.73} 
\definecolor{mediumcandyapplered}{rgb}{0.89, 0.02, 0.17}

\usepackage[most]{tcolorbox}
\tcbuselibrary{breakable}
\tcbset{every box/.style={enhanced,breakable}}
\tcbset{colframe=black,colback=red!10,enhanced,breakable,sharp corners}

\usepackage{enumitem} 

\usepackage{notation}

\usepackage{adjustbox}

\definecolor{alex}{RGB}{51,183,150}
\definecolor{erik}{RGB}{235,134,52}
\definecolor{Vicky}{RGB}{235,134,52}

\newcommand{\ticked}{$\text{\rlap{$\checkmark$}}\square$}
\newcommand{\unticked}{{$\square$}}
\newcommand{\tick}[1]{\ifthenelse{#1=1}{\ticked}{\unticked}}

\newcommand{\rmv}{\hspace*{-.3mm}}

\newcommand{\vm}[1]{\ensuremath{\bm{#1}}} 

\renewcommand{\exp}[1]{\ensuremath{{e}^{#1}}}

\renewcommand{\Re}[1]{\ensuremath{\text{Re}\!\left[#1\right]}}
\renewcommand{\Im}[1]{\ensuremath{\text{Im}\!\left[#1\right]}}

\newcommand{\norm}[2]{\ensuremath{\lVert #1 \rVert^{#2}}}

\newcommand{\transp}{\ensuremath{^\mathsf{T}}}

\newcommand{\minus}{\rmv - \rmv}

\newcommand{\s}{\hspace*{0.5pt}}

\providecommand{\norm}[1]{\lVert#1\rVert}

\newcommand{\ist}{\hspace*{.3mm}}
\newcommand{\iist}{\hspace*{1mm}}

\newcommand{\nn}{\nonumber}

\mathchardef\Re="023C
\mathchardef\Im="023D

\newlength{\figureheight}
\newlength{\figurewidth}
\graphicspath{{./figures/}}

\allowdisplaybreaks

\definecolor{mycolor01}{rgb}{0.00000,0.00000,1.00000}
\definecolor{mycolor02}{rgb}{0.133,0.545,0.133}
\definecolor{mycolor03}{rgb}{0.50000,0.00000,0.50000}
\definecolor{mycolor05}{rgb}{1.00000,0.83984,0.00000}
\definecolor{mycolor04}{rgb}{0.92969,0.50781,0.92969}
\definecolor{mycolor06}{rgb}{1.00000,0.64453,0.00000}
\definecolor{mycolor07}{rgb}{0.50000,0.50000,0.50000}
\definecolor{mycolor08}{rgb}{1.00000,0.00000,0.00000}
\definecolor{mycolor09}{rgb}{0.2510 ,0.8784, 0.8157}
\definecolor{mycolor10}{rgb}{0.54297,0.00000,0.00000}
\definecolor{mycolor11}{rgb}{0.6445, 0.1641,0.1641}
\definecolor{mycolor12}{rgb}{1, 0, 1}
\definecolor{APDA}{rgb}{0.4, 0.75, 0.5}
\definecolor{APPDA}{rgb}{0.1, 0.6, 0.3}
\definecolor{APPDA2}{rgb}{0.0, 0.5, 0.2}
\definecolor{APROP}{rgb}{0.396, 0.663, 0.843}
\definecolor{APPROP}{rgb}{0.549, 0.647, 0.918}
\definecolor{plum}{rgb}{0.537,0.243,0.506}

\pgfdeclarelayer{back1}
\pgfdeclarelayer{lay1}
\pgfdeclarelayer{back2}
\pgfdeclarelayer{lay2}
\pgfsetlayers{back2,lay2,back1,main,lay1}

\makeatletter

\tikzset{
	nomorepostactions/.code={\let\tikz@postactions=\pgfutil@empty},
	decmark/.style 2 args={decoration={markings,
			mark= between positions 0 and 1 step (1/6)*\pgfdecoratedpathlength with{%
				\tikzset{#2,every mark}\tikz@options
				\pgftransformresetnontranslations
				\pgfuseplotmark{#1}%
			},  
		},
		postaction={decorate},
		/pgfplots/legend image post style={
			mark=#1, mark options={#2}, every path/.append style={nomorepostactions}
		},
	},
	markbeginend/.style 2 args={decoration={markings,
			mark= between positions 0 and 1 step (1)*\pgfdecoratedpathlength with{%
				\tikzset{#2,every mark}\tikz@options
				\pgfuseplotmark{#1}%
			},  
		},
		postaction={decorate},
		/pgfplots/legend image post style={
			mark=#1,mark options={#2},every path/.append style={nomorepostactions}
		},
	},
	markend/.style 2 args={decoration={markings,
			mark= at position \pgfdecoratedpathlength with{%
				\tikzset{#2,every mark}\tikz@options
				\pgfuseplotmark{#1}%
			},  
		},
		postaction={decorate},
		/pgfplots/legend image post style={
			mark=#1,mark options={#2},every path/.append style={nomorepostactions}
		},
	},
	posmark/.style 2 args={decoration={markings,
			mark= at position #2 with{%
				\tikzset{solid,every mark}\tikz@options
				\pgftransformresetnontranslations
				\pgfuseplotmark{#1}%
			},  
		},
		postaction={decorate},
		/pgfplots/legend image post style={
			mark=#1,mark options={solid},every path/.append style={nomorepostactions}
		},
	},
}

\makeatother

\pgfplotsset{
	resultStyle1/.style={mark=none, line width=0.5pt, mycolor01, decmark={oplus}{solid}},
	resultStyle2/.style={mark=none, line width=0.5pt, mycolor02, decmark={triangle}{solid}},
resultStyle3/.style={mark=none ,line width=0.5pt, mycolor03, decmark={+}{solid}},
resultStyle4/.style={mark=none, line width=0.5pt, mycolor06, decmark={star}{solid}},
resultStyle5/.style={mark=none, line width=0.5pt, mycolor08, decmark={o}{solid}},
resultStyle6/.style={mark=none, line width=0.5pt, mycolor05, decmark={square}{solid}}, 
resultStyle7/.style={mark=none, line width=0.5pt, mycolor09, decmark={diamond}{solid}}, 
resultStyle8/.style={mark=none, line width=0.5pt, mycolor11, decmark={otimes}{solid}}, 
resultStyle9/.style={mark=none, line width=0.5pt, mycolor12, decmark={x}{solid}}, 
resultStyleBase/.style={mark=none, line width=0.5pt,}, 
compareStyle1/.style={mark=none, line width=0.5pt, mycolor01},
compareStyle2/.style={mark=none, line width=0.5pt, mycolor02},
compareStyle3/.style={mark=none ,line width=0.5pt, mycolor03},
compareStyle4/.style={mark=none, line width=0.5pt, mycolor06},
compareStyle5/.style={mark=none, line width=0.5pt, mycolor08},
compareStyle6/.style={mark=none, line width=0.5pt, mycolor05}, 
compareStyle7/.style={mark=none, line width=0.5pt, mycolor09}, 
compareStyle8/.style={mark=none, line width=0.5pt, mycolor11}, 
compareStyle9/.style={mark=none, line width=0.5pt, mycolor12}, 
}

\pgfplotsset{
compat=newest,
simple style group/.style={
label style={font=\scriptsize},
legend style={font=\scriptsize},
tick label style={font=\scriptsize},
nodes near coords style={font=\scriptsize},
title style={font=\scriptsize},
scale only axis,
grid style={dotted},
mark options={solid}, 
},
simple style/.style={
label style={font=\scriptsize},
legend style={font=\scriptsize},
tick label style={font=\scriptsize},
nodes near coords style={font=\scriptsize},
title style={font=\scriptsize},
width=\figurewidth,
height=\figureheight,
at={(0\figurewidth,0\figureheight)},
scale only axis,
grid style={dotted},
mark options={solid}, 
},
base style/.style={
label style={font=\scriptsize},
legend style={font=\scriptsize},
tick label style={font=\scriptsize},
nodes near coords style={font=\scriptsize},
title style={font=\scriptsize},
width=\figurewidth,
height=\figureheight,
at={(0\figurewidth,0\figureheight)},
scale only axis,
cycle list={
	{mark=none, line width=0.5pt, mycolor01, solid},
	{mark=none, line width=0.5pt, mycolor02, dash dot},
	{mark=none ,line width=0.5pt, mycolor03, densely dashed},
	{mark=none, line width=0.5pt, mycolor04, dash dot dot},
	{mark=x   , line width=0.5pt, mycolor05},
	{mark=.   , line width=0.7pt, mycolor06}, 
	{mark=square,only marks, mark size = 0.8pt, mycolor07,
		mark options = {line width = 0.4pt}},
	{mark=x,     only marks, mark size = 1.3pt, mycolor08,
		mark options = {line width = 0.4pt}},
	{mark=o,     only marks, mark size = 0.8pt, mycolor09,
		mark options = {line width = 0.4pt}},
	{mark=o, mycolor10},
},
grid style={dotted},
xmajorgrids,
ymajorgrids,
mark options={solid}, 
},
base style group/.style={
	label style={font=\scriptsize},
	legend style={font=\scriptsize},
	tick label style={font=\scriptsize},
	nodes near coords style={font=\scriptsize},
	title style={font=\scriptsize},
	scale only axis,
	grid style={dotted},
	xmajorgrids,
	ymajorgrids,
	mark options={solid}, 
},
std graph style new/.style={
xlabel style={yshift=1mm},
ylabel style={yshift=-1.5mm},
yticklabel style={xshift=1mm},
},
color lines style/.style={
cycle list={
	{mark=none, mycolor01, decmark={oplus}{solid} },
	{mark=none, mycolor02, decmark={+}{solid} }, 
	{mark=none, mycolor03, decmark={triangle}{solid} }, 
	{mark=none, mycolor04, decmark={star}{solid} }, 
	{mark=none, mycolor05, decmark={o}{solid} },
	{mark=none, mycolor06, decmark={square}{solid} },
},
},
meas graph style/.style={
xlabel style={yshift=1mm},
ylabel style={yshift=-1mm},
xmajorgrids,
ymajorgrids,
mark repeat = 1,
mark phase = 0,
cycle list={
	{color=black, only marks, mark=*, mark size=0.5pt, mark options={solid, black}},
	{color=red, only marks, mark=*, mark size=0.1pt, line width=0.25pt},
},
ylabel={},
}, 
ci graph style/.style={
xlabel style={yshift=1mm},
ylabel style={yshift=-1.5mm},
yticklabel style={xshift=1mm},
mark repeat = 1,
mark phase = 0,
ymin=1e-3,
ymax=100,
ytick = {100, 50, 10, 1, 0.1, 0.01, 1e-3, 1e-4},
yticklabels = {$0$, $50$, $90$, $99$, $99.9$, $99.99$, $99.999$, $99.9999$},
y dir=reverse,
},     
bp coeff style/.style={
scale only axis=true,
width=0.225*.9\linewidth,
height=0.225*.9\linewidth,
scale only axis,
xmin=-4.000,
xmax=4.000,
xlabel={$\ell${\color{white}$\aod$}},
ticklabel style={font=\footnotesize},
ymin=0.000, ymax=0.9,
ylabel={$c_\ell$},
xlabel style={font=\footnotesize},
ylabel style={font=\footnotesize},
major tick length=2pt
},
bp graph style/.style={        
scale only axis=true,
width=0.35*1.1\linewidth,
height=0.225*.9\linewidth,
scale only axis,
xmin=-3.14, xmax=3.14,
xlabel={$\aod${\color{white}$\ell$}},
ticklabel style={font=\footnotesize},
xtick={-3.14,-1.57,0.0,1.57,3.14},
xticklabels={$-\pi$,$-\tfrac{\pi}{2}$,$0$,$\tfrac{\pi}{2}$,$\pi$},
ymin=0.000, ymax=3,
ylabel={Beampattern},
xlabel style={font=\footnotesize}, ylabel style={font=\footnotesize},
major tick length=2pt
},
peb graph style/.style={        
width=0.66\linewidth,
scale only axis,
point meta min=-2.583,
point meta max=-0.300,
axis on top,
xmin=0.000,
xmax=12.000,
xlabel={x in meter},
y dir=reverse,
ymin=0.000,
ymax=8.000,
ylabel={y in meter},
ytick={7.0,6.0,...,0.0},
xtick={0.0,1.0,...,12.0},
yticklabels={$1$,$2$,$3$,$4$,$5$,$6$,$7$,$8$},
xlabel style={font=\scriptsize,yshift=0.125cm},
ylabel style={font=\scriptsize,yshift=-0.125cm},
ticklabel style={font=\scriptsize},
unit vector ratio*=1 1 1,
yticklabel pos=left,
major tick length=2pt,
colormap={mymap}{[1pt] rgb(0pt)=(1,1,1); rgb(1pt)=(0.858903,0.984776,0.839302); rgb(2pt)=(0.777958,0.94143,0.649487); rgb(3pt)=(0.755504,0.864264,0.463393); rgb(4pt)=(0.777509,0.754439,0.310168); rgb(5pt)=(0.820314,0.619497,0.21003); rgb(6pt)=(0.854796,0.471879,0.170327); rgb(7pt)=(0.851327,0.326629,0.183322); rgb(8pt)=(0.784671,0.198575,0.225774); rgb(9pt)=(0.637629,0.0993149,0.259577); rgb(10pt)=(0.400067,0.0343393,0.229819); rgb(11pt)=(0,0,0)},
colorbar style={ylabel={Position Error Bound in centimeter (logscale)}, ytick={-0.4,-0.82,...,-2.92}, yticklabels={$39.8$, $15.1$, $5.8$, $2.2$, $0.8$, $0.3$},ylabel style={yshift=0.5mm,font=\scriptsize,scale=0.8},width=2.0mm,xshift=-4.25mm,ticklabel style={font=\scriptsize},major tick length=0pt}, 
colormap access=piecewise constant
},
peb ellipses/.style={color=white, line width=0.4pt, forget plot}
}

\tikzset{naming/.style={align=center,font=\small}}
\tikzset{antenna/.style={insert path={-- coordinate (ant#1) ++(0,0.25) -- +(135:0.25) + (0,0) -- +(45:0.25)}}}
\tikzset{station/.style={naming,draw,shape=dart,shape border rotate=90, minimum width=10mm, minimum height=10mm,outer sep=0pt,inner sep=3pt}}
\tikzset{mobile/.style={naming,draw,shape=rectangle,minimum width=12mm,minimum height=6mm, outer sep=0pt,inner sep=3pt}}
\tikzset{radiation/.style={{decorate,decoration={expanding waves,angle=90,segment length=4pt}}}}

\tikzset{
  pobl/.style={
    inner sep=0pt, outer sep=0pt, fill=#1,
  },
  pobl gron/.style n args={2}{
    pobl=#1, rounded corners=#2,
  },
  pics/person/.style n args={3}{
    code={
      \node (-corff) [pobl=#1, minimum width=.25*#2, minimum height=.375*#2, rotate=#3, pic actions] {};
      \node (-pen) [minimum width=.3*#2, circle, pobl=#1, outer sep=.01*#2, anchor=south, rotate=#3, pic actions] at (-corff.north) {};
      \node (-coes dde) [pobl gron={#1}{1pt}, anchor=north west, minimum width=.12125*#2, minimum height=.25*#2, rotate=#3, pic actions] at (-corff.south west) {};
      \node [pobl=#1, anchor=north, minimum width=.12125*#2, minimum height=.15*#2, rotate=#3, pic actions] at (-coes dde.north) {};
      \node (-coes chwith) [pobl gron={#1}{1pt}, anchor=north east, minimum width=.12125*#2, minimum height=.25*#2, rotate=#3, pic actions] at (-corff.south east) {};
      \node [pobl=#1, anchor=north, minimum width=.12125*#2, minimum height=.15*#2, rotate=#3, pic actions] at (-coes chwith.north) {};
      \node (-braich dde) [pobl gron={#1}{.75pt}, minimum width=.075*#2, minimum height=.325*#2, outer sep=.0064*#2, anchor=north west, rotate=#3, pic actions] at (-corff.north east)  {};
      \node [pobl=#1, minimum width=.05*#2, minimum height=.2*#2, outer sep=.0064*#2, anchor=north west, rotate=#3, pic actions] at (-corff.north east) {};
      \node (-braich chwith) [pobl gron={#1}{.75pt}, minimum width=.075*#2, minimum height=.325*#2, outer sep=.0064*#2, anchor=north east, rotate=#3, pic actions] at (-corff.north west) {};
      \node [pobl=#1, minimum width=.0375*#2, minimum height=.2*#2, outer sep=.0064*#2, anchor=north east, rotate=#3, pic actions] at (-corff.north west) {};
      \node (-fit person) [fit={(-pen.north) (-braich dde.east) (-coes chwith.south) (-braich chwith.west)}] {};
    },
  },
  pics/SBS/.style={code={
      \begin{scope}[local bounding box=#1]
      \fill [pic actions/.try] (-1,0) -- (-1/2,3) -- (1/2, 3) -- (1,0) -- cycle;
      \fill [pic actions/.try] (-1/16,2) rectangle (1/16,4);
      \fill [pic actions/.try] (0,4) circle [radius=1/4];
      \foreach \i in {-1,1}
        \fill [shift=(90:4), xscale=\i]
          \foreach \r in {1,3/2,2}{
            (-45:\r) arc (-45:45:\r) -- (45:\r-1/10)
            arc(45:-45:\r-1/10) -- cycle
          };
       \end{scope}
  }},
}

\begin{document}

\title{Multi-Sensor Fusion for Extended Object Tracking Exploiting Active and Passive Radio Signals
}

\author{Hong Zhu$^{1,2}$,~\IEEEmembership{Student Member,~IEEE}, Alexander Venus$^{1,2}$,~\IEEEmembership{Member,~IEEE}, \\Erik Leitinger$^{1,2}$,~\IEEEmembership{Member,~IEEE} and Klaus Witrisal$^{1,2}$, ~\IEEEmembership{Member,~IEEE}
	\thanks{The financial support by the Christian Doppler Research Association, the Austrian Federal Ministry for Digital and Economic Affairs and the National Foundation for Research, Technology and Development is gratefully acknowledged.}	
	\\
	\small{{$^1$Graz University of Technology, Austria},
	}\\
	\small{{$^2$Christian Doppler Laboratory for Location-aware Electronic Systems}}\\
}

\maketitle
\frenchspacing

\renewcommand{\baselinestretch}{1}\small\normalsize 

\begin{abstract}
	Reliable and robust positioning of radio devices remains a challenging task due to multipath propagation, hardware impairments, and interference from other radio transmitters. A frequently overlooked but critical factor is the agent itself, e.g., the user carrying the device, who potentially obstructs line-of-sight (LOS) links to the base stations (anchors). This paper addresses the problem of accurate positioning in scenarios where LOS links are partially blocked by the agent. The agent is modeled as an extended object (EO) that scatters, attenuates, and blocks radio signals. We propose a Bayesian method that fuses ``active'' measurements (between device and anchors) with ``passive'' multistatic radar-type measurements (between anchors, reflected by the EO). To handle measurement origin uncertainty, we introduce an multi-sensor and multiple-measurement probabilistic data association (PDA) algorithm that jointly fuses all EO-related measurements. Furthermore, we develop an EO model tailored to agents such as human users, accounting for multiple reflections scattered off the body surface, and propose a simplified variant for low-complexity implementation. Evaluation on both synthetic and real radio measurements demonstrates that the proposed algorithm outperforms conventional PDA methods based on point target assumptions, particularly during and after obstructed line-of-sight (OLOS) conditions.

\end{abstract}

\acresetall 

\begin{IEEEkeywords} extended object tracking, robust radio positioning, obstructed line-of-sight (OLOS), passive radar measurements,  probabilistic data association (PDA) \end{IEEEkeywords}

\IEEEpeerreviewmaketitle



\section{Introduction}\label{sec:introduction}

Radio localization has witnessed significant advancements in recent years, playing a crucial role in autonomous navigation, indoor positioning, and smart healthcare. It determines an object's position by analyzing properties of received radio signals, such as time of arrival, received signal strength, or angle of arrival, to infer distances or angles relative to known reference points. 
Radar sensing has been extensively utilized for object detection and motion tracking by analyzing the reflection, Doppler shift, and scattering properties of reflected signals~\cite{Wymeersch2022, WitrisalSPM2016Copy}. 
Traditional monostatic and bistatic radar systems detect objects by transmitting signals and receiving their reflections using co-located or separated antennas~\cite{Gedschold2023, Zetik2007}. The integration of radar sensing with radio localization enables simultaneous positioning and tracking, which is crucial for applications such as vehicular networks, robotics, and human activity recognition \cite{PonteMueller2023}.
Additionally, multi-sensor methods have emerged as a robust approach for integrating data from spatially distributed agents and for leveraging multipath propagation to support localization~\cite{ZhaStaJosWanGenDamWymHoeTAES2020,  Mendrzik2019}. 
These methods exploit the knowledge of the geometric propagation environment, spatial diversity, and sensor fusion techniques to enhance localization accuracy and robustness in complex, dynamic environments~\cite{LeiVenTeaMey:TSP2023}. However, when the radio device is carried by an agent, the human body can introduce significant impairments by blocking, attenuating, or scattering the signals, thereby reducing the reliability of direct \ac{los} measurements~\cite{Ambroziak2016}.
\begin{figure}[t]
	\centering
	\includegraphics[width=9.5cm]{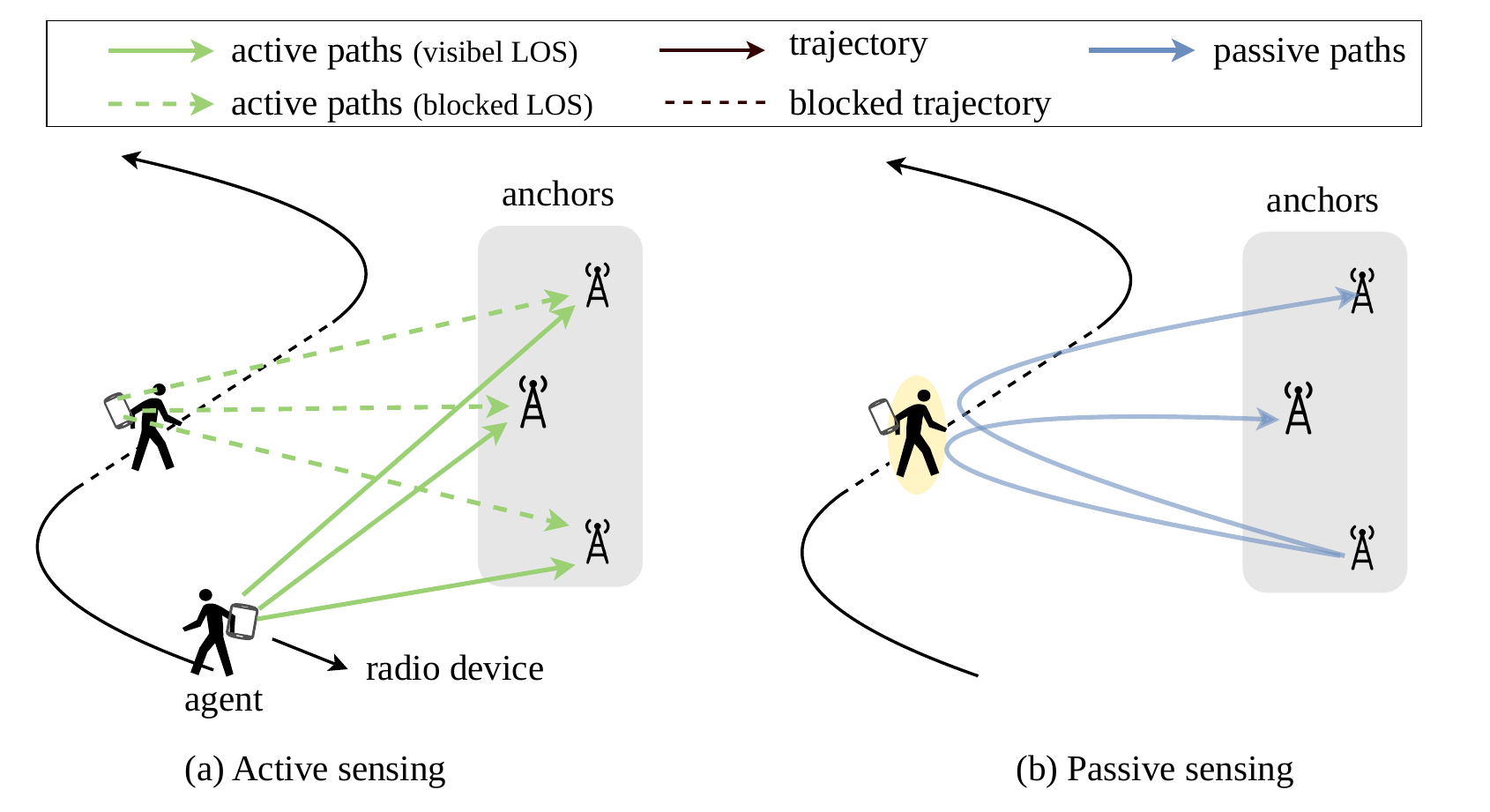}
	\vspace{-6mm}
	\caption{{Overview of the considered sensing scenario. The agent carrying a radio device moves along a trajectory. (a) Active sensing with LOS paths to anchors. (b) Passive sensing via scatter propagation under OLOS conditions, where LOS paths are blocked by the agent itself. The agent is modeled as an EO that blocks, scatters and attenuates radio signals.}
}
	\label{fig:concept}
	\vspace{-6mm}
\end{figure}

\subsection{State-of-the-Art Methods}\label{sec:referenceMethods}

In recent years, the concept of \ac{eot} has emerged as a powerful paradigm for scenarios in which objects give rise to multiple scattering paths due to their physical extent \cite{GraBau:JAIF2017}.
Unlike traditional point-target tracking methods that model objects as dimensionless entities, \ac{eot} explicitly incorporates the spatial extent of the object, enabling more accurate modeling of such scenarios.
Two prominent approaches to \ac{eot} are star-convex shape models~\cite{Baum2014, Hirscher2016, Kumru2021} and the random matrix framework. The latter represents the spatial distribution of scattering points using parameterized geometric shapes, most commonly ellipses~\cite{Koch2008, Feldmann2011, Schuster2015, Zhang2021, Hoher2022} and rectangles~\cite{Granstroem2011, Granstroem2014}, embedded in a probabilistic state estimation framework.
{The random matrix approach is widely used in the \ac{eot} problem, which uses a traditional elliptical \ac{eo} model  assuming scatter points are uniformly distributed on the full object volume~\cite{Koch2008}.
In addition, ~\cite{Hoher2022} introduces a directional-support assumption for elliptical extended objects, where only the portions within the sensor’s field of view can generate measurements. However, this formulation is based on a single lidar sensor.}
\Ac{eot} methods have been successfully applied in tracking vehicles, human motion, and other extended objects~\cite{Schuster2015, Kumru2021}. 

\Ac{pda} is a Bayesian approach widely used in target tracking to address measurement origin uncertainty. It leverages extracted measurement information, such as time delays and amplitudes, to update the state estimation of an object in the presence of clutter\cite{BarShalomTCS2009}. Traditional \ac{pda} methods assume the presence of a single object-related measurement per sensor (point object assumption) and disregard the extended nature of objects \cite{JeoTugTAES2005, Venus2021}. This approach fails in scenarios where multiple measurements are generated from spatially distributed scattering points, leading to reduced localization performance.
A solution is to incorporate multiple object-related measurements into the data association process\cite{MeyerTSP2021, Wielandner2024}. 
However, as the number of sensors in the network grows, so does the complexity of the underlying algorithms required for data fusion and object tracking. In this context, there is a critical need for scalable and computationally efficient models that can process information from diverse sensors without compromising performance or localization accuracy.

The performance of active radio localization (of a transmitting radio device) can degrade severely under \ac{olos} conditions. For example, when a radio device is carried by an agent (e.g. a person or a robot), the \ac{los} radio links can be obstructed by the agent itself during certain time.
To address this  problem, off-body channel properties have been studied, which describe  interactions of radio signals transmitted between a wearable
radio device and some external infrastructure ~\cite{Ambroziak2016, WildingPIMRC2020}. Incorporating this insight in tracking algorithms has the potential to enhance the robustness and accuracy of radio localization systems, particularly in challenging environments where multipath effects and signal obstructions are prevalent.

\subsection{Contribution}\label{sec:contribution}

In this work, we present a Bayesian method that fuses \emph{active} measurements between the radio device mounted on the agent and the anchors, as well as \emph{passive}, radar-like measurements between the anchors where signals are reflected off the agent. The algorithm is tailored to \ac{olos} scenarios, in which the \ac{los} of active measurements is potentially obstructed by the agent itself (see Fig.~\ref{fig:concept}). The agent, which blocks, scatters, and attenuates radio signals, is modeled as an \ac{eo}. The aim is to jointly estimate the position of the radio device and the position and extent of the \ac{eo}. 
We use the geometric information embedded in the active measurements as well as in the passive measurements. 
Since passive measurements do not originate from the full object volume, we propose an \ac{eo} model of the agent that considers the fact that measurements only originate from the surface and that each anchor-anchor pair observes the measurements from different directions.
The sequential estimation of the device and \ac{eo} states is achieved by computing the marginal posterior \ac{pdf} through passing messages on a factor graph according to the \ac{spa}, using a particle-based implementation.
The contributions of this paper are as follows.
\begin{itemize}
	\item We introduce an \ac{eo} model that accounts for \acp{mpc} scattered off the agent's surface.
	 {In contrast to the traditional elliptical EO model, the proposed formulation explicitly captures direction-dependent visibility and surface scattering behavior.}
	
	\item We develop a {\emph{simplified} \ac{eo} model} that captures the scattering behavior of an extended object using simplified geometric shapes, thereby reducing the computational complexity.
	
	\item We derive a likelihood model that enables the fusion of active measurements (between a radio device and anchors), along with passive measurements (resulting from the surface scattering at the agent).

	\item We propose a multi-sensor, multiple-measurement \ac{pda} algorithm that efficiently fuses data of all \ac{eo}-related measurements from all anchors.

\end{itemize}
This paper advances over the preliminary account of our method provided in the conference publications \cite{Zhu2024, Zhu2025} by 
	{(i) a detailed derivation of the \ac{eo} model and simplified \ac{eo} models, including \ac{ut}-based likelihood derivation; (ii) a complete Bayesian factor-graph formulation with full \ac{spa} message updates and particle-based implementation; (iii) an evaluation of the efficiency of the proposed EO model with different anchor geometries; (iv) a performance comparison between the proposed algorithm and the traditional EO-based algorithm; (v) an analysis of time-varying extent estimation with proposed \ac{eo} models in \ac{olos} scenarios; 
	(vi) an expanded evaluation using measurements extracted from simulated radio signals at multiple bandwidths; and (vii) an evaluation of the proposed algorithms using real measured data, and a table to summarize the averaged runtime and \ac{rmse} of compared methods.  }

\subsection{Notation}
We define the following \acp{pdf} with respect to $\mathbf{x} \in \mathbb{R}^n$ or $\rv{x} \in \mathbb{R}$. 
The $n$-dimensional Gaussian \ac{pdf}  with mean $\boldsymbol{\mu} \in \mathbb{R}^n$ and covariance matrix $\boldsymbol{\Sigma} \in \mathbb{R}^{n \times n}$ \cite{Kay1998} is defined as 
\begin{align}
	f_\mathrm{N}(\mathbf{x}; \boldsymbol{\mu}, \boldsymbol{\Sigma}) = 
	\frac{1}{(2\pi)^{n/2} \, |\boldsymbol{\Sigma}|^{1/2}} 
	\exp{\left( -\frac{1}{2} (\mathbf{x} - \boldsymbol{\mu})^\top \boldsymbol{\Sigma}^{-1} (\mathbf{x} - \boldsymbol{\mu}) \right)}.
	\label{eq:nd_gaussian_pdf}
\end{align}
Let $\mathcal{S} \subset \mathbb{R}^n$ be a bounded, measurable support set with volume $|\mathcal{S}|$. The $n$-dimensional uniform \ac{pdf} over $\mathcal{S}$ is defined as
\begin{align}
	\mathcal{U}(\mathbf{x}; \mathcal{S}) = \frac{1}{|\mathcal{S}|} \, \mathbf{1}_{\mathcal{S}}(\mathbf{x}).
	\label{eq:2unipdf}\\[-7mm]\nn
\end{align}
Finally, the Gamma \ac{pdf} is defined as 
\vspace*{-1mm}
\begin{align}
	\mathcal{G}(x;\alpha,\beta) =& \frac{1}{\beta^\alpha \Gamma(\alpha)} x^{\alpha-1} e^{-\frac{x}{\beta}}
	\label{eq:gammapdf}\\[-7mm]\nn
\end{align}
where $\alpha$ is the shape parameter, $\beta$ is the scale parameter and $\Gamma(\cdot)$ is the gamma-function.




 
\section{System Model and Problem Formulation}\label{sec:system_model}

\subsection{Problem Formulation}\label{sec:problem}

\begin{table*}[bt] \vspace*{-3mm}
	\renewcommand{\baselinestretch}{1}\small\normalsize
	\setlength{\tabcolsep}{3pt} %
	\renewcommand{\arraystretch}{1.1} %
	\footnotesize
	\centering
	\caption{Summary and description of parameters of the system models.}\label{tbl:rvs}
	\begin{tabular}{r|ccc|ccc|cc|c|cc}
		\toprule
		\textbf{Model Type} & \multicolumn{3}{c|}{\textbf{Kinematic state} $\V{x}_n$} &\multicolumn{3}{c|}{\textbf{Extent state} $\V{X}_n$} & \multicolumn{2}{c|}{\textbf{Deterministic variables}}& \textbf{Constant} &\multicolumn{2}{c}{\textbf{Scatter variables}}\\ \midrule
		\textbf{EO Model} & position & velocity & bias & \makecell{semi-major\\axis} & \makecell{semi-minor\\axis} & \makecell{annular\\width}  & FoV start & FoV end &{} & active & passive\\
		\textbf{} & $\RV{p}_n$ & $\RV{v}_n$ & $\RV{b}_n$ & ${{l}_n}$ & ${{w}_n}$ & ${\delta_n}$  & $\varphi_{s \s n}^{(j,j')}/\varphi_{s \s n}^{(j)}$ & $\varphi_{e \s n}^{(j,j')}/\varphi_{e \s n}^{(j)}$ & {} & $\V{q}_{\text{A},n}^{(j)}$ &  $\V{q}_{\text{P},n}^{(j,j')}$\\ \midrule
		\makecell{\textbf{Simplified} \\ \textbf{EO model}}  & position & velocity & bias & \multicolumn{2}{c}{{radius}} & \makecell{sca. ellipse\\semi-minor axis} & \makecell{sca. ellipse\\semi-major axis} &\makecell{sca. ellipse\\orientation}& {opening angle} & active & passive\\
		\textbf{} & $\RV{p}_n$ & $\RV{v}_n$ & $\RV{b}_n$ & \multicolumn{2}{c}{$r_n$} & $w_{s \s n}$ &   $l_{s\s n}$  & $\theta_{s\s n}^{(j)}$&  {{$\alpha$}} & $\V{\zeta}_{n}^{(j)}$ & $\V{\zeta}_{n}^{(j)}$ \\ 
		\bottomrule
	\end{tabular}
	\vspace{-2.5mm}
\end{table*}

\begin{figure}[t]
	\centering
	\captionsetup[subfloat]{captionskip=-2mm}
	\subfloat[\label{fig:ModelidealActive}]{\includegraphics{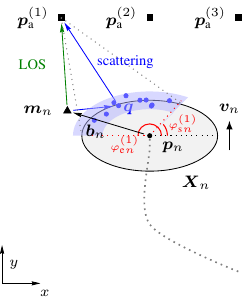}}\vspace{0mm}
	\subfloat[\label{fig:ModelidealPassive}]{\includegraphics{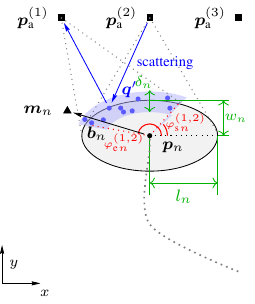}}\vspace{0mm}
	\caption{Illustration of the \emph{\ac{eo} model} observed by three anchors at time $n$ with exemplary signal propagation paths for (a) active measurements from the radio device $\V{m}_n$ to the receiving anchor $j=1$, and (b) passive measurements between transmitting anchor $t=2$ and receiving anchor $j=1$.}
	\label{fig:Modelideal}
	\vspace{-6mm}
\end{figure}

The aim fo this paper is to estimate the position of a radio device $\V{m}_n \in \mathbb{R}^2$, mounted on an agent moving with velocity $\V{v}_n = [v_{\text{x}\s n}\; v_{\text{y}\s n}]^\text{T}$, {where $n$ is the time index}. The agent is modeled as an \ac{eo} with an extent state $\V{X}_n$. The \ac{eo} is centered at position $\V{p}_n = [p_{\text{x}\s n} \iist p_{\text{y}\s n} ]^\text{T}$, and is rigidly coupled to the radio device. The bias vector between the radio device and the \ac{eo} center is denoted by $\V{b}_n = [b_{\rho\s n} \iist b_{\phi\s n}]^\text{T}$, where $b_{\rho\s n}$ is the distance from the \ac{eo} center to the device, and $b_{\phi\s n}$ is the orientation angle relative to the x-axis of the \ac{eo} coordinate system. 
At time $n$, the extended object, and the radio device coupled to the EO, are described by the kinematic state $\V{x}_n = [\V{p}_n ^\text{T}\; \V{v}_n ^\text{T}\; \V{b}_n^\text{T}]^\text{T}$ and extent state  $\V{X}_n$. 
The orientation of the \ac{eo} is decided as $\theta_n = \text{atan}(v_{\text{y}\s n}/v_{\text{x}\s n})$.
The radio device position is thus given by $\V{m}_{n} = \V{p}_n+ [b_{\rho \s n} \cos(b_{\phi \s n}+\theta_n) \iist\iist b_{\rho \s n} \sin(b_{\phi \s n}+\theta_n)]\transp$. Consequently, estimating $\V{p}_n$, $\V{v}_n$, and $\V{b}_n$ (together with $\V{X}_n$) is sufficient to determine $\V{m}_n$.

The device transmits an \emph{active} radio signal at time $n$. Each anchor $j \in \{1,\s... \s, J \}$ at known position $\V{p}_{\text{a}}^{(j)} = [p_{\text{ax}}^{(j)} \iist p_{\text{ay}}^{(j)}]^\text{T}$, serves as a receiver. Simultaneously, anchors $j' \in \mathcal{A}_\text{tx}$ transmit \emph{passive} radio signals, which are received by all anchors $j \in \{1,\s... \s, J\}$. 
{In the active signal model, the dominant geometric factor is the direct LOS propagation between the radio device and each anchor, with additional EO-induced scattering paths, whereas in the passive signal model, all informative components arise purely from anchor–EO–anchor bistatic scattering paths.}
The \emph{active} and \emph{passive} radio signals are jointly used to estimate the kinematic state $\V{x}_n$ and the extent state $\V{X}_n$, making it possible to estimate the radio device position $\V{m}_n$ even if the device is in \ac{olos} to all anchors. Details about the radio signal models can be found in Appendix~\ref{sec:RSmodel}.

\subsection{EO Models}\label{sec:EOmodels}

We introduce two models describing the \ac{eo}, referred to as the \emph{\ac{eo} model} and the \emph{simplified \ac{eo} model}. Both are described by their spatial states $\V{p}_n$,  $\V{v}_n$, $\V{b}_n$, and extent state $\V{X}_n$, as summarized in Table~\ref{tbl:rvs}. Note that we use the same symbol for the extent state vector, $\V{X}_n$, in both the \ac{eo} model and the simplified \ac{eo} model, and explicitly indicate the model when a distinction is necessary.
{Although the physical size of the extended object is assumedly constant, the extent parameters $\bm{X}_n$ are modelled as time-dependent states in the Bayesian
	filter, because their true values are unknown and must be inferred and gradually
	refined over time during the sequential estimation process.}

\subsubsection{EO Model}\label{sec:ideal_concept}

Previous studies have shown that scatter points are primarily located along the surface of the \ac{eo}, exhibiting a truncated \ac{fov}, as described in~\cite{Hoher2022, WildingPIMRC2020}. As illustrated in Fig.~\ref{fig:Modelideal}, the proposed \ac{eo} scattering model assumes that scatter points originate from an annular sector on the \ac{eo}'s surface, where they are uniformly distributed. The \ac{eo} is modeled as an ellipse with semi-major axis $l_n$, semi-minor axis $w_n$, and radial width $\delta_n$ of the annular sector. The extent state is defined as {$\V{X}_n = [l_n\;\; w_n\;\; \delta_n]^\text{T}$}.

For \emph{active measurements}, the scattering sector is defined by the individual \ac{fov} of the receiving anchor $j$ at position $\V{p}_a^{(j)}$, as shown in Fig.~\ref{fig:ModelidealActive}. This sector is defined as the arc on the \ac{eo} bounded by two tangents from the anchor position, resulting in an angular extent bounded by $\varphi_{s \s n}^{(j)}$ and $\varphi_{e \s n}^{(j)}$. The corresponding scatter point
${\bm{q}} \triangleq \V{q}_{\text{A},n}^{(j)}$ 
is assumed to be uniformly distributed within this sector, i.e., $f(\V{q}_{\text{A},n}^{(j)}|\V{X}_n, \V{x}_n) = \mathcal{U}(\V{q}_{\text{A},n}^{(j)}; \mathcal{S}^{(j)}(\V{X}_n,\V{x}_n))$, where $\mathcal{S}^{(j)}(\V{X}_n,\V{x}_n)$ denotes the support region defined by the individual \ac{fov} of anchor $j$ and the extent parameters $\V{X}_n$, while the position and orientation of the \ac{eo} is given by the kinematic state $\V{x}_n$.

For \emph{passive measurements}, the angular span of the scattering sector is determined by the \emph{common \ac{fov}} between a transmitting anchor $j'$ at position $\V{p}_a^{(j')}$ and a receiving anchor $j$ at position $\V{p}_a^{(j)}$, as depicted in Fig.~\ref{fig:ModelidealPassive}. Each anchor’s FoV is defined as the arc on the \ac{eo} bounded by the tangents from its location. The intersection of these arcs defines the common FoV of the anchor pair $(j,j')$, whose angular extent, bounded by $\varphi_{s \s n}^{(j,j')}$ and $\varphi_{e \s n}^{(j,j')}$, specifies the relevant scattering sector on the \ac{eo}. The scatter point 
${\bm{q'}} \triangleq \V{q}_{\text{A},n}^{(j,j')}$
is assumed to be uniformly distributed within this sector, i.e., $f(\V{q}_{\text{P},n}^{(j,j')} | \V{X}_n,\V{x}_n) = \mathcal{U}(\V{q}_{\text{P},n}^{(j,j')}; \mathcal{S}^{(j,j')}(\V{X}_n,\V{x}_n))$, where $\mathcal{S}^{(j,j')}(\V{X}_n,\V{x}_n)$ denotes the support region defined  by the common \ac{fov} of anchor pair $(j,j')$ and the extent parameters $\V{X}_n$, while the position and orientation of the \ac{eo} is given by the kinematic state $\V{x}_n$.

\subsubsection{Simplified EO Model} \label{sec:approximate_concept}

\begin{figure}[t]
	\centering
	\captionsetup[subfloat]{captionskip=-2mm}
	\subfloat[\label{fig:ModelEOActive}]{\includegraphics{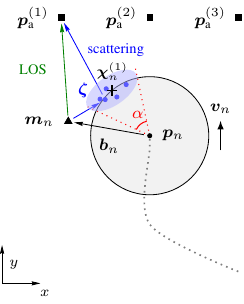}}\vspace{0mm}
	\subfloat[\label{fig:ModelEOPassive}]{\includegraphics{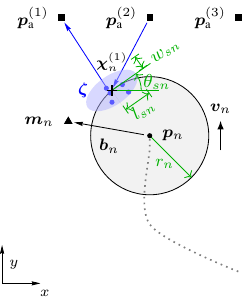}}\vspace{0mm}
	\caption{Illustration of the {\emph{simplified \ac{eo} model}} observed by three anchors at time $n$ with exemplary signal propagation paths for (a) active measurements from the radio device $\V{m}_n$ to the receiving anchor $j=1$, and (b) passive measurements between transmitting anchor $t=2$ and receiving anchor $j=1$.}
	\label{fig:EOModel_2case}
\end{figure}

To reduce the computational complexity while preserving key geometric features, the simplified \ac{eo} model provides an efficient alternative to the \ac{eo} model in Section~\ref{sec:ideal_concept}. It approximates the spatial scattering behavior illustrated in Fig.~\ref{fig:Modelideal} using a geometric representation, illustrated by the blue ellipse  in Fig.~\ref{fig:EOModel_2case}. In this model, the \ac{eo} is simplified to a circular shape with radius $r_n$, and the scattering behavior is captured by an anchor-dependent scattering ellipse on the surface from which the measurements originate.

As shown in Fig.~\ref{fig:EOModel_2case}, each scattering ellipse is centered at a point $\V{\chi}_n^{(j)}$ located on the boundary of the circular \ac{eo}, and its shape is defined by a semi-major axis {$l_{\text{s}\s n}$} and a semi-minor axis $w_{\text{s}\s n}$. 
{The semi-major axis is determined by a constant opening angle $\alpha$ around the line connecting the \ac{eo} center $\V{p}_n$ to the position of the $j$th anchor $\V{p}_{\text{a}}^{(j)}$. 
Since radius $r_n$ of the circular EO is updated over time, the resulting semi-major axis $l_{s \s n}$ is anchor-independent but varies with the time index $n$.
The semi-minor axis $w_{\text{s}\s n}$ is assumed to be identical for all receiving anchors at each time $n$.} The extent state of the {simplified EO model} is thus defined as $\V{X}_n = [r_n \ist w_{\text{s}\s n}]^\text{T}$.

For each receiving anchor $j$, the center of the corresponding scattering ellipse is computed as
\begin{equation}
	\V{\chi}_n^{(j)} = \V{p}_n +
	\begin{bmatrix}
		r_n \cos(\phi_n^{(j)}) \\
		r_n \sin(\phi_n^{(j)})
	\end{bmatrix}
	\label{equ:intersection}
\end{equation}
where $\phi_n^{(j)} = \text{atan}\big(\big(p_{\text{y}\s n} - p_{\text{a}\s \text{y}}^{(j)}\big)/\big(p_{\text{x}\s n} - p_{\text{a}\s \text{x}}^{(j)}\big)\big)$ denotes the angle from the receiving anchor to the center of the \ac{eo}. The orientation of the scattering ellipse is given by $\theta_{\text{s}\s n}^{(j)} = \pi - \phi_n^{(j)}$, which corresponds to the tangent direction of the circle at the scatter point.

The scattering ellipse is represented by the positive semi-definite matrix {$\V{R}_n^{(j)} = \V{A}_n^{(j)} \V{E}_n \V{A}_n^{(j)\text{T}} \in \mathbb{R}^{2\times2}$}, referred to as the scattering matrix,
where $\V{A}_n^{(j)}$ is a rotation matrix defined as
\begin{equation}
	\V{A}_n^{(j)} =
	\begin{bmatrix}
		\cos(\theta_{\text{s}\s n}^{(j)}) & -\sin(\theta_{\text{s}\s n}^{(j)}) \\
		\sin(\theta_{\text{s}\s n}^{(j)}) & \cos(\theta_{\text{s}\s n}^{(j)})
	\end{bmatrix}
\end{equation}
and $\V{E}_n^{(j)}$ is a diagonal matrix containing the squared half-axes of the ellipse
{\begin{equation}
		\V{E}_n =
		\begin{bmatrix}
			e_{1\s n} & 0 \\
			0 & e_{2\s n}
		\end{bmatrix}
\end{equation}}
with eigenvalues {$e_{1\s n} = \big(l_{\text{s}\s n}/2\big)^2$} and {$e_{2\s n} = \big(w_{\text{s}\s n}/2\big)^2$}.

The scatter points ${\bm{\zeta}} \triangleq \V{\zeta}_{n}^{(j)}$
associated with the ellipse of the receiving anchor $j$ are then modeled as being drawn from a Gaussian \ac{pdf} $f(\V{\zeta}_n^{(j)}|\V{X}_n, \V{x}_n) \triangleq f_{\mathrm{N}}(\V{\zeta}_n^{(j)}; \V{\chi}_n^{(j)}, \V{R}_n^{(j)})$ with mean $\V{\chi}_n^{(j)}$ and covariance $\V{R}_n^{(j)}$. This probabilistic representation also captures the spatial spread of scatter points on the \ac{eo} surface. 
{While $\V{\zeta}_{n}^{(j)}$ is drawn from the same spatial distribution for both active and passive cases, the specific realizations of each case are independent.}
{We note that the simplified EO model is not a direct probabilistic approximation of the EO model; instead, it provides a simplified circle model in which the locally visible surface region is represented by an anchor-dependent Gaussian scattering ellipse.}

\subsection{State Transition Model} \label{sec:SSM}
To simplify the derivation, we define the augmented extended object state as $\V{y}_n = [{\V{x}}_n^\text{T}\; \V{X}_n ^\text{T} ]^\text{T}$.
It is assumed that the states $\V{y}_n$ evolve over time $n$ as independent first-order Markov processes. Therefore, the joint state transition \ac{pdf} can be represented as 
\begin{align}
	f(\V{y}_n | \V{y}_{n-1}) 
	= f(\V{x}_{n}|\V{x}_{n-1}) f(\V{X}_{n}|\V{X}_{n-1})
	\label{equ:state_transition_model}
\end{align}
where $f({\V{x}}_n|{\V{x}}_{n-1})$ and $f(\V{X}_n|\V{X}_{n-1})$ are the state transition \acp{pdf} of the kinematic and extent states.
The kinematic state $\V{x}_n$ contains both motion vectors (position $\V{p}_n$ and velocity $\V{v}_n$) and the bias vector $\V{b}_n$. 
We assume they evolve independently. Therefore, the kinematic transition \ac{pdf} is factorized as:
$f(\V{x}_n | \V{x}_{n-1}) = f(\V{p}_n, \V{v}_n | \V{p}_{n-1}, \V{v}_{n-1}) f(\V{b}_n | \V{b}_{n-1})$.
The agent motion, described by $f(\V{p}_n, \V{v}_n | \V{p}_{n-1}, \V{v}_{n-1})$, follows a linear, constant velocity and stochastic acceleration model, given as $[\V{p}_n^\text{T} \; \V{v}_n^\text{T}]^\text{T} = \V{A}\, [\V{p}_{n\minus 1}^\text{T} \; \V{v}_{n\minus 1}^\text{T}]^\text{T} + \V{B}\, \RV{w}_{n}$. The acceleration process $\RV{w}_n$ is i.i.d. across $n$, zero mean, and Gaussian with covariance matrix ${\sigma_{\text{a}}^2}\, \V{I}_2$, with ${\sigma_{\text{a}}}$ being the acceleration standard deviation, and $\V{A} \in \mathbb{R}^{\text{4x4}}$ and $\V{B} \in \mathbb{R}^{\text{4x2}}$ are defined according to \cite[p.~273]{BarShalom2002EstimationTracking}.
The PDF of the bias state $\V{b}_n$ is factorized as $f({b}_n|{b}_{n-1})= f({b}_{\rho \s n}|{{b}}_{\rho \s {n-1}})f({b}_{\phi \s n}|{b}_{\phi \s {n-1}})$.
The state transition PDF of the angular parameter ${b}_{\phi \s n}$ is modeled by the wrapped Gaussian distribution $f({b}_{\phi \s n}|{b}_{\phi \s {n-1}}) = f_{\text{WN}}({b}_{\phi \s n}; {b}_{\phi \s {n-1}}, \sigma_{\phi}) = \sum_{k=-\infty}^{\infty} f_{\text{N}}({b}_{\phi \s n}+2\pi k; {b}_{\phi \s {n-1}}, \sigma_{\phi}) $ with the fact that ${b}_{\phi \s n} \in [-\pi,\pi)$. 
{The wrapped Gaussian distribution is selected for angular variables to inherently follow circular periodicity and avoid boundary discontinuities.} 
The length parameter ${b}_{\rho \s n}$ is strictly positive; therefore, its state-transition PDF is modeled by a Gamma distribution given by  $f({b}_{\rho \s n}|{b}_{\rho \s {n-1}}) = \mathcal{G}( {b}_{{\rho} \s {n}}; \kappa_{\rho}, {b}_{\rho \s {n-1}}/ \kappa_{\rho})$. 
{This parametrization ensures $\mathbb{E}[{b}_{\rho \s n}] = {b}_{\rho \s {n-1}}$, thus preserving the constant-length assumption in the absence of observations. The variance is $\mathrm{Var}({b}_{\rho \s n}) = {b}_{\rho \s {n-1}}^2/\kappa_{\rho}$, implying that the variance scales with the object size. The parameter $\kappa_{\rho}$ controls the state noise, where smaller values correspond to larger deviations from the previous state.}
Furthermore, the PDF of the extent state $\V{X}_n$ is discussed in two cases:
\subsubsection{EO Model}
The PDF of the extent state $\V{X}_n$ is factorized as $f(\V{X}_n|\V{X}_{n-1})= f({l}_{n}|{{l}}_{{n-1}})f({w}_{n}|{{w}}_{{n-1}}) f({\delta}_{n}|{\delta}_{{n-1}})$, and assumed to evolve independently of each other across time $n$. 
The dynamics of the individual state variables are modeled by Gamma \acp{pdf}  $f(l_n|{l}_{n-1}) = \mathcal{G}({l}_{n}; \kappa_{\text{l}}, {l}_{n-1}/ \kappa_{\text{l}})$, $f(w_n|w_{n-1}) = \mathcal{G}({w}_{n}; \kappa_{\text{w}}, {w}_{n-1}/ \kappa_{\text{w}})$, and $f(\delta_n|\delta_{n-1}) = \mathcal{G}({\delta}_{n}; \kappa_{\delta}, {\delta}_{n-1}/ \kappa_{\delta})$ \footnote{{We use independent Gamma distributions to model the extent parameters. This choice is consistent with standard extended object models employing positive-support distributions (e.g., Wishart-distributed extent matrices \cite{Koch2008}), while accommodating the surface-scattering assumption adopted in this work.}}

\subsubsection{Simplified EO Model}
The PDF of the extent state $\V{X}_n$ is factorized as $f(\V{X}_n|\V{X}_{n-1})= f({r}_{n}|{{r}}_{{n-1}})f(w_{\text{s}\s n}|w_{\text{s}\s {n-1}})$, and assumed to evolve independently of each other across time $n$. 
The dynamics of the individual state variables are modeled by Gamma \acp{pdf}  $f({r}_{n}|{{r}}_{{n-1}}) = \mathcal{G}({r}_{n}; \kappa_{\text{r}}, {r}_{n-1}/ \kappa_{\text{r}})$ and $f({w}_{\text{s}\s n}|{{w}}_{\text{s}\s {n-1}}) = \mathcal{G}({w}_{\text{s}\s n}; \kappa_{\text{w}_\text{s}}, {w}_{\text{s}\s n-1}/ \kappa_{\text{w}_\text{s}})$.

\subsection{Measurement \ac{lhf} Model} \label{sec:measurement_model}
\emph{Measurement Extraction:} At time $n$, for each received anchor $j$ and each anchor-anchor pair $(j,j')$, measurements are extracted from the active radio and passive radio signals (see Appendix~\ref{sec:RSmodel}) using a \ac{ceda} \cite{Hansen2018, Grebien2024}. {Active and passive measurements are collected simultaneously and processed separately during the measurement campaign. Therefore the \ac{ceda} is performed separately for the two measurement sets.} The vector containing active measurements is defined as $\V{z}^{(j)}_{\text{A},n}= [\V{z}^{(j)}_{\text{A},n,1},\dots\,, \V{z}^{(j)}_{\text{A},n, M_{\text{A}, n}^{(j)}}]$ with $M_{\text{A},n}^{(j)}$ being the number of active measurements. Each active measurement $\V{z}^{(j)}_{\text{A},n,l} = [z_{\text{A},\text{d},n,l}^{(j)} ~ z_{\text{A},\text{u},n,l}^{(j)}]^\text{T}$, $l \in  \{1,\,\dots\,M_{\text{A},n}^{(j)}\}$ contains a distance measurement $z_{\text{A},\text{d},n,l} \in [0, d_\text{max}]$, where $d_\text{max}$ is the maximum measurement distance, and a normalized amplitude measurement  $z_{\text{A}, \text{u},n,l} \in [\gamma, \infty)$ corresponding to the square root of the estimated component \ac{snr}, where $\gamma$ is the detection threshold. The vector containing passive measurements is defined as $\V{z}^{(j,j')}_{\text{P},n}= [\V{z}^{(j,j')}_{\text{P},n,1},\dots\,, \V{z}^{(j,j')}_{\text{P},n, M_{\text{P}, n}^{(j,j')}}]$ with $M_{\text{P},n}^{(j,j')}$ being the number of passive measurements. Each passive measurement $\V{z}^{(j,j')}_{\text{P},n,l}= [z_{\text{P},\text{d},n,l}^{(j,j')} ~ z_{\text{P},\text{u},n,l}^{(j,j')}]^\text{T}$, $l \in  \{1,\,\dots\,M_{\text{P},n}^{(j,j')}\}$ contains a distance measurement $z_{\text{P},\text{d},n,l}^{(j,j')} \in [0, d_\text{max}]$ and a normalized amplitude measurement $z_{\text{P}, \text{u},n,l}^{(j,j')} \in [\gamma, \infty)$. 

{In our statistical measurement model, we assume that all entries of $\V{z}^{(j)}_{\text{A},n}$ and $\V{z}^{(j,j')}_{\text{P},n}$ are conditionally independent given the states $x_n$ and $X_n$.}
Furthermore, the measurements are assumed to be statistically independent across time steps $n$, anchors $j$, and anchor pairs $(j,j')$, where $j \in \{1,\dots,J\}$ and $j' \in \mathcal{A}_\text{tx}$. The \ac{lhf} for the individual active and passive measurements are introduced in what follows.

\subsubsection{Active Measurement \ac{lhf}}\label{sec:measLikelihood_active}

The active measurements can originate from either a direct path measurement or \ac{eo}-related scattering measurements and the according \ac{lhf} is given by
\begin{align}
	&\hspace*{-2mm}f_\text{A}(\V{z}^{(j)}_{\text{A},n,l}|\V{x}_n,\V{X}_n) \nn\\
	&\hspace*{-2mm}=P_{\text{mix}}f_{\text{D}}(\V{z}^{(j)}_{\text{A},n,l}|\V{x}_n)+ (1-P_{\text{mix}}) 
	f_{\text{AS}}(\V{z}^{(j)}_{\text{A}, n,l}|\V{x}_n,\V{X}_n)
	\label{equ:measLikelihood_active}
\end{align}
where $P_{\text{mix}}$ represents the probability that the measurement originates from the direct path. 
{The active measurements may contain both LOS and EO-related scattering components simultaneously. However, these are modeled as independent measurement contributions in the likelihood function, and their presence depends on the propagation and visibility conditions.}
The \ac{lhf} of the direct path is given by
\begin{align} 
	&f_{\text{D}}(\V{z}^{(j)}_{\text{A},n,l}|\V{x}_n) = f_{\text{N}}(z_{\text{A},\text{d},n,l}^{(j)}; h^{(j)}(\V{x}_n), \sigma_{\text{A},n,l}^{(j)\ist 2}) 
	\label{equ:measLikelihood_active_LOS}
\end{align}
where $f_{\text{N}}(x;\mu,\sigma^2)$ is the Gaussian PDF, with mean $h^{(j)}(\V{x}_n) = \|\vm{m}_n - \vm{p}_\text{a}^{(j)}\|$ 
being the LOS measurement function 
(recall that $\V{m}_{n} = \V{p}_n+ [b_{\rho \s n} \cos(b_{\phi \s n}+\theta_n) \iist\iist b_{\rho \s n} \sin(b_{\phi \s n}+\theta_n)]\transp$ is calculated using $\V{x}_n$), and variance $\sigma_{\text{A},n,l}^{(j)\ist 2} \triangleq \sigma^{2}(z^{(j)}_{\text{A},\text{u},n,l})$. The variance is determined from the Fisher information given by
$ \sigma^{2}(z^{(j)}_{\text{A},\text{u},n,l}) =   c^2 / ( 8\,  \pi^2 \, \beta_\text{bw}^2 \, z^{(j)\ist 2}_{\text{A},\text{u},n,l})$, where $\beta_\text{bw}$ is the root mean squared bandwidth \cite{WitrisalJWCOML2016,LeitingerJSAC2015}.
The \ac{lhf} of the EO-related scattering paths $f_{\text{AS}}(\V{z}^{(j)}_{\text{A}, n,l}|\V{x}_n,\V{X}_n)$ is described in Sec.~\ref{sec:scattering_model}.

\subsubsection{Passive Measurement \ac{lhf}}\label{sec:measLikelihood_passive}

The passive measurements \ac{lhf} $f_{\text{PS}}(\V{z}^{(j,j')}_{\text{P}, n,l}|\V{x}_n,\V{X}_n)$ originate only from \ac{eo}-related scattering paths.

\subsection{Single Scattering Path \ac{lhf} Model}\label{sec:scattering_model}

\subsubsection{\ac{eo} \ac{lhf}}\label{sec:ideal_model}
 
We consider the \ac{eo} model introduced in Section~\ref{sec:ideal_concept}. The corresponding \ac{eo} \acp{lhf} for the active and passive cases are obtained by marginalizing out the scatter point variables $\V{q}_{\text{A},n}^{(j)}$ and $\V{q}_{\text{P},n}^{(j,j')}$ \cite{Koch2008}, respectively, resulting in 
\begin{align}
	&f_{\text{AS}}(\V{z}^{(j)}_{\text{A}, n,l}|\V{x}_n, \V{{X}}_n)  \nn \\ 
	&\hspace*{10mm}=\int f(\V{z}^{(j)}_{\text{A},n,l}|{\V{x}}_{n}, \V{q}_{\text{A},n}^{(j)})f(\V{q}_{\text{A},n}^{(j)}|\V{{X}}_{n}, \V{x}_n) d\V{q}_{\text{A},n}^{(j)}
	\label{equ:measLikelihood_idealAS} \\
	&f_{\text{PS}}(\V{z}^{(j,j')}_{\text{P}, n,l}|\V{x}_n, \V{{X}}_n)  \nn \\ 
	&\hspace*{10mm}=\int f(\V{z}^{(j,j')}_{\text{P},n,l}|\V{x}_{n}, \V{q}_{\text{P},n}^{(j,j')})f(\V{q}_{\text{P},n}^{(j,j')}|\V{{X}}_{n}, \V{x}_n) d\V{q}_{\text{P},n}^{(j,j')}
	\label{equ:measLikelihood_idealPS} 
\end{align}

{The integrals in \eqref{equ:measLikelihood_idealAS} and \eqref{equ:measLikelihood_idealPS} consist of two components: (i) the Gaussian distribution for measurement noise
$f(\V{z}^{(j)}_{\text{A},n,l}|\V{x}_n, \V{q}_{\text{A},n}^{(j)}) = f_{\text{N}}(z_{\text{A},\text{d},n,l}^{(j)}; h^{(j)}_\text{A}(\V{m}_n,\V{q}_{\text{A},n}^{(j)}), \sigma_{\text{A},n,l}^{(j)\ist 2})$  and $f(\V{z}^{(j,j')}_{\text{P},n,l}|\V{x}_{n}, \V{q}_{\text{P},n}^{(j,j')}) = f_{\text{N}}(z_{\text{P},\text{d},n,l}^{(j,j')}; h^{(j,j')}_\text{P}(\V{q}_{\text{P},n}^{(j,j')}), \sigma_{\text{P},n,l}^{(j)\ist 2})$, whose means are given by the nonlinear distances 
$h^{(j)}_\text{A}(\V{m}_n,\V{q}_{\text{A},n}^{(j)}) =\|\V{q}_{\text{A},n}^{(j)} - \V{m}_n\|+ \|\V{q}_{\text{A},n}^{(j)} - \vm{p}_\text{a}^{(j)}\|$ and 
$h^{(j,j')}_\text{P}(\V{q}_{\text{P},n}^{(j,j')})= \|\V{q}_{\text{P},n}^{(j,j')} - \V{p}_\text{a}^{(t)}\|+ \| \V{q}_{\text{P},n}^{(j,j')} - \V{p}_\text{a}^{(j)}\|$, and
(ii) the scatter point distribution 
$f(\V{q}_{\text{A},n}^{(j)}|\V{{X}}_{n}, \V{x}_n)$ and $f(\V{q}_{\text{P},n}^{(j,j')}|\V{{X}}_{n}, \V{x}_n)$, which are uniform over an elliptic annular sector on the EO surface.
These integrals do not admit a closed-form solution because the Gaussian means depend nonlinearly on the integration variables $\V{q}_{\text{A},n}^{(j)}$ and $\V{q}_{\text{P},n}^{(j,j')}$,
while the uniform distribution is defined over an elliptic annular sector with curved, nonlinear boundaries. As a result, the marginal likelihoods become analytically intractable integrals.}

\emph{Approximation of the integrals:} The integrals in \eqref{equ:measLikelihood_idealAS} and \eqref{equ:measLikelihood_idealPS} cannot be solved analytically. Therefore, we resort to Monte Carlo integration by means of importance sampling \cite{DouWan:SPM2005} yielding
\vspace*{-2mm}
\begin{align}
	&f_{\text{AS}}(\V{z}^{(j)}_{\text{A}, n,l}|\V{x}_n, \V{{X}}_n) \nn\\
	&\hspace*{8mm}\approx \frac{1}{I} \sum_{i=1 }^{I} f_{\text{N}}(z_{\text{A},\text{d},n,l}^{(j)};h^{(j)}_\text{A}(\V{m}_n, \V{q}^{(j)[i]}_{\text{A},n}), \sigma_{\text{A},n,l}^{(j)\ist 2}) 
	\label{equ:measLikelihood_idealAS_appr}\\
	&f_{\text{PS}}(\V{z}^{(j,j')}_{\text{P}, n,l}|\V{x}_n, \V{{X}}_n) \nn\\
	&\hspace*{8mm} \approx \frac{1}{I} \sum_{i=1 }^{I} f_{\text{N}}(z_{\text{P},\text{d},n,l}^{(j,j')}; h^{(j,j')}_\text{P}( \V{q}^{(j,j')[i]}_{\text{P},n}), \sigma_{\text{P},n,l}^{(j,j')\ist 2}) 
	\label{equ:measLikelihood_idealPS_appr}\\[-8mm]\nn
\end{align}
where $\sigma_{\text{P},n,l}^{(j,j')\ist 2} \triangleq \sigma^{2}(z^{(j,j')}_{\text{P},\text{u},n,l})$, $\V{q}^{(j)[i]}_{\text{A}, n}$ and $\V{q}^{(j,j')[i]}_{\text{P}, n}$ are the samples drawn from the importance distribution and $I$ is the number of samples. Details about how to draw samples from the importance distribution can be found in Appendix \ref{scatterPDF}.

\subsubsection{Simplified \ac{eo} \ac{lhf}} \label{sec:approximate_model}

Similar to \eqref{equ:measLikelihood_idealAS} and \eqref{equ:measLikelihood_idealPS}, the approximate \acp{lhf} for the active and passive scatter points are obtained by marginalizing out the corresponding scatter point random variables $\V{\zeta}_{n}^{(j)}$ \cite{Koch2008}, resulting in
\begin{align}
	&f_{\text{AS}}(\V{z}^{(j)}_{\text{A}, n,l}|\V{x}_n, \V{X}_n)  \nn \\
	&\hspace*{10mm}=\int f(\V{z}^{(j)}_{\text{A},n,l}|\V{x}_n,\V{\zeta}_{n}^{(j)}) f(\V{\zeta}_n^{(j)}|\V{X}_n, \V{x}_n) d\V{\zeta}_{n}^{(j)}
	\label{equ:measLikelihood_AS}\\
	&f_{\text{PS}}(\V{z}^{(j,j')}_{\text{P},n,l}|\V{x}_n, \V{X}_n) \nn\\
	&\hspace*{10mm}=\int f(\V{z}^{(j,j')}_{\text{P},n,l}|\V{x}_n, \V{\zeta}_{n}^{(j)})f(\V{\zeta}_{n}^{(j)}|\V{X}_n, \V{x}_n) d\V{\zeta}_{n}^{(j)}
	\label{equ:measLikelihood_PS}
\end{align}
where $f(\V{z}^{(j)}_{\text{A},n,l}|\V{x}_n, \V{\zeta}_{n}^{(j)}) = f_{\text{N}}(z_{\text{A},\text{d},n,l}^{(j)}; h^{(j)}_\text{A}(\V{m}_n,\V{\zeta}_{n}^{(j)}), \sigma_{\text{A},n,l}^{(j)\ist 2})$, $f(\V{z}^{(j)}_{\text{P},n,l}|\V{x}_n, \V{\zeta}_{n}^{(j)}) = f_{\text{N}}(z_{\text{P},\text{d},n,l}^{(j,j')}; h^{(j,j')}_\text{P}(\V{\zeta}_{n}^{(j)}), \sigma_{\text{P},n,l}^{(j,j')\ist 2})$ and $f(\V{\zeta}_n^{(j)}|\V{X}_n, \V{x}_n) \triangleq f_{\mathrm{N}}(\V{\zeta}_n^{(j)}; \V{\chi}_n^{(j)}, \V{R}_n^{(j)})$ as described in Sec.~\ref{sec:approximate_concept}

\emph{Approximation of the integrals:} The integrals in \eqref{equ:measLikelihood_AS} and \eqref{equ:measLikelihood_PS} are not analytically tractable. However, since all involved \acp{pdf} are Gaussian and the nonlinear functions exhibit only weak nonlinearity within the support of the Gaussian  prior, we approximate the integrals using the \ac{ut} \cite{Ristic2003}. In particular, the \acp{sp} for $\V{\zeta}_{n}^{(j)} \in \mathbb{R}^2$ are determined by the covariance matrix $\V{R}^{(j)}_n \in \mathbb{R}^{2\times2}$, which parameterizes the \acp{pdf} $f_{\mathrm{N}}(\V{\zeta}_{n}^{(j)}; \V{\chi}^{(j)}_n, \V{R}_n^{(j)})$. This results in a total of $2D+1$ \acp{sp} for $D = 2$, given by \cite[Chapter 2.4]{Ristic2003}
\vspace*{-1mm}
\begin{align}
	\V{\zeta}_{n}^{(j)\ist[d]} \rmv\rmv=\rmv\rmv \begin{cases}
		\V{\chi}^{(j)}_n, \rmv\rmv&\text{$d = 0$}\\ 
		\V{\chi}^{(j)}_n \rmv\rmv\rmv+\rmv\rmv\rmv \sqrt{D+\kappa}[\V{R}_n^{(j)\ist \frac{1}{2}}]_{d}, &\text{$d \rmv\rmv=\rmv\rmv 1, \dots, 2$}\\
		\V{\chi}^{(j)}_n \rmv\rmv\rmv-\rmv\rmv\rmv\sqrt{D+\kappa}[\V{R}_n^{(j)\ist \frac{1}{2}}]_{d-D}, &\text{$d \rmv\rmv=\rmv\rmv 3, \dots, 4$}
	\end{cases}
	\label{equ:calcSP}\\[-6mm]\nn
\end{align}
with according weights
\vspace*{-1mm}
\begin{align}
	w^{[d]} = \begin{cases}
		\frac{\kappa}{2+\kappa}, &\text{$d = 0$}\\ 
		\frac{1}{2(2+\kappa)}, &\text{$d = 1, \dots, 4$}
	\end{cases}
	\label{equ:SPweights}
\end{align}
where $[\cdot]_{d}$ denotes the $d$th column of a matrix, and $\kappa$ is a parameter that controls the spread of the \acp{sp} with respect to the mean value of the transformed random variable \cite[Chapter 2.4]{Ristic2003}. Note that, since the covariance matrix $\V{R}^{(j)}_n$ is identical for active and passive measurements, only one set of \acp{sp} $\V{\zeta}_{n}^{(j)\ist[d]}$ is used. Applying the \ac{ut} using \eqref{equ:calcSP} and \eqref{equ:SPweights}, 
the first and second moments of the transformed \acp{sp} are given as
\vspace*{-1mm}
\begin{align}
	\mu^{(j)}_{\text{A},n} &= \sum_{d=0}^{2D} w^{(d)} h^{(j)}_\text{A}(\V{m}_n, \V{\zeta}_{n}^{(j)\ist[d]}) \label{equ:transMeanA}\\
	\mu^{(j,j')}_{\text{P},n} &= \sum_{d=0}^{2D} w^{(d)}  h^{(j,j')}_\text{P}(\V{\zeta}_{n}^{(j)\ist[d]})\label{equ:transMeanP}\\
	\sigma_{\text{A},n}^{(j)\ist 2} &= \sum_{d=0}^{2D} w^{(d)} (h^{(j)}_\text{A}(\V{m}_n, \V{\zeta}_{n}^{(j)\ist[d]})- \mu^{(j)}_{\text{A},n} )^2\label{equ:transCovaA}\\
	\sigma_{\text{P},n}^{(j,j')\ist 2} &= \sum_{d=0}^{2D} w^{(d)} ( h^{(j,j')}_\text{P}(\V{\zeta}_{n}^{(j)\ist[d]}) - \mu^{(j,j')}_{\text{P},n} )^2\label{equ:transCovaP}\ist.\\[-6mm]\nn
\end{align}
Using \eqref{equ:transMeanA} and \eqref{equ:transCovaA}, as well as \eqref{equ:transMeanP} and \eqref{equ:transCovaP}, the \ac{ut}-based approximations of the \ac{eo} \acp{lhf} in \eqref{equ:measLikelihood_AS} and \eqref{equ:measLikelihood_PS} are then given by
\begin{align}
	\hspace*{-2mm}&f_{\text{AS}}(\V{z}^{(j)}_{\text{A}, n,l}|\V{x}_n, \V{X}_n) \rmv\rmv=\rmv\rmv f_{\text{N}}(z_{\text{A},\text{d},n,l}^{(j)}; \mu^{(j)}_{\text{A},n}, \sigma_{\text{A},n}^{(j)\ist 2} + \sigma_{\text{A},n,l}^{(j)\ist 2})
	\label{equ:measLikelihood_ASApprox}\\
	\hspace*{-2mm}&f_{\text{PS}}(\V{z}^{(j,j')}_{\text{P},n,l}|\V{x}_n, \V{X}_n) \rmv\rmv=\rmv\rmv f_{\text{N}}(z_{\text{P}, \text{d},n,l}^{(j,j')}; \mu^{(j,j')}_{\text{P},n}\rmv\rmv\rmv\rmv\rmv, \sigma_{\text{P},n}^{(j,j')\ist 2} \rmv\rmv+\rmv\rmv\sigma_{\text{P},n,l}^{(j,j')\ist 2}) \ist.
	\label{equ:measLikelihood_PSApprox}
\end{align}

\subsection{Data Association Uncertainty}\label{sec:dataAssociation}

For each anchor $j$ and anchor pair $(j,j')$, the measurements $\V{z}_{\text{A},n}^{(j)}$ and $\V{z}_{\text{P},n}^{(j,j')}$ contain both object-related components and false positives.
It is not known whether a measurement originates from the device or the \ac{eo}. It is also not known if a measurement does not originate from either of these, i.e., if it is a false positive. 
{The false positive corresponds to spurious components estimated (e.g., amplitude and delay statistics) produced by the CEDA algorithm, which may result from noise, interference, sidelobes of the ambiguity function, or reflections from surrounding structures.} 
The association variables $a_{\text{A},n,l}^{(j)} \in \{0,1\}$ and $a_{\text{P},n,l}^{(j,j')} \in \{0,1\}$  specify whether a single measurement is object-related, which is denoted by $1$, or not, which is denoted by $0$. 
{The \ac{pda} algorithm assigns a soft association probability to each detected MPC, reflecting how likely it is to be EO-related rather than a false positive. This enables the filter to incorporate all available measurements in a weighted manner, without making hard association decisions.}

{The EO surface offers many potential scatter points, each with a low probability of producing a detectable MPC. Therefore, 
the number of EO-related measurements is assumed to follow a Poisson distribution with mean $\mu_{\text{m}}(\V{x}_n,\V{X}_n)$}\footnote{The number of EO-related measurements is influenced by the size and orientation of the EO\cite{MeyerTSP2021}.}. For both passive and active cases, the number of false positive measurements is also modeled as Poisson distributed with mean $\mu_{\text{fp}}$. These false positive measurements are assumed to be independent and distributed according to uniform PDFs $f_\text{fp}(\V{z}_{\text{A},n,l}^{(j)})$ and $f_\text{fp}(\V{z}_{\text{P},n,l}^{(j,j')})$, respectively.
{The uniform PDFs reflect the assumption that such a false positive can occur at any delay within the detectable range $[0, d_\text{max}]$, as long as it exceeds the detection threshold $\gamma$. Consequently, we define $f_\text{fp}(\cdot) = \frac{1}{d_\text{max}}$ for $z_{\text{d},n,l} \in [0, d_\text{max}]$, and zero otherwise.}

{The factors relating the prior distributions of the number of measurements and false positives to the association vectors of active and passive measurements, $\V{a}^{(j)}_{\text{A},n}= [a_{\text{A},n,1}^{(j)} \ist \cdots \ist a_{\text{A},n,M^{(j)}_{\text{A},n}}^{(j)}]\transp$ and $\V{a}^{(j,j')}_{\text{P},n}= [a_{\text{P},n,1}^{(j,j')} \ist \cdots \ist a_{\text{P},n,M^{(j,j')}_{\text{P},n}}^{(j,j')}]\transp$, is given as~\cite[supplementary material]{MeyerTSP2021}}
\begin{align}
	\eta_\text{A}(a_{\text{A},n,l}^{(j)}) &= \begin{cases}
		\frac{\mu_{\text{m}}(\V{x}_n, \V{X}_n)}{\mu_{\text{fp}}}, &\text{$a_{\text{A},n,l}^{(j)} = 1$}\\ 
		1, & \text{$a_{\text{A},n,l}^{(j)} = 0$}
	\end{cases}
	\label{equ:PMFassocVarA}\\
	\eta_\text{P}(a_{\text{P},n,l}^{(j,j')}) &= \begin{cases}
	\frac{\mu_{\text{m}}(\V{x}_n, \V{X}_n)}{\mu_{\text{fp}}}, &\text{$a_{\text{P},n,l}^{(j,j')} = 1$}\\ 
	1, & \text{$a_{\text{P},n,l}^{(j,j')} = 0$}
	\end{cases}\ist.
	\label{equ:PMFassocVarP}
\end{align}
We denote by $\V{a}_{\text{A},{1:n}} = [\V{a}_{\text{A},1}^\text{T}\, \dots\, \V{a}_{\text{A},n}^\text{T}]^\text{T}$ with $\V{a}_{\text{A},n'} = [\V{a}^{(1)\ist\text{T}}_{\text{A},n'}\ist\cdots\ist \V{a}^{(J)\ist\text{T}}_{\text{A},n'}]\transp$ and $\V{a}_{\text{P},{1:n}} = [\V{a}_{\text{P},1}^\text{T}\, \dots\, \V{a}_{\text{P},n}^\text{T}]^\text{T}$ with $\V{a}_{\text{P},n'} =  [\V{a}_{\text{P},n'}^{(1,1)\ist\text{T}}\ist\cdots\ist\V{a}_{\text{P},n'}^{(1,|\mathcal{A}_\text{tx}|)\ist\text{T}}\ist\cdots\ist\V{a}_{\text{P},n'}^{(J,|\mathcal{A}_\text{tx}|)\ist\text{T}}]\transp$.

\subsection{Global \ac{lhf}}

Since we assume that all entries of $\V{z}^{(j)}_{\text{A},n}$ and $\V{z}^{(j,j')}_{\text{P},n}$ are {conditionally independent} and that the measurements are statistically independent across time steps $n$, anchors $j$, and anchor pairs $(j,j')$, where $j \in \{1,\dots,J\}$ and $j' \in \mathcal{A}_\text{tx}$.
We define 
$\V{z}_{\text{A},1:n} = [\V{z}_{\text{A},1}^\text{T}\, \dots\, \V{z}_{\text{A},n}^\text{T}]^\text{T}$ with  $\V{z}_{\text{A},n'} = [\V{z}_{\text{A},n'}^{(1)\ist\text{T}}\ist\cdots\ist\V{z}_{\text{A},n'}^{(J)\ist\text{T}}]\transp$ and $\V{z}_{\text{P},1:n} = [\V{z}_{\text{P},1}^\text{T}\, \dots\, \V{z}_{\text{P},n}^\text{T}]^\text{T}$ with $\V{z}_{\text{P},n'} =  [\V{z}_{\text{P},n'}^{(1,1)\ist\text{T}}\ist\cdots\ist\V{z}_{\text{P},n'}^{(1,|\mathcal{A}_\text{tx}|)\ist\text{T}}\ist\cdots\ist\V{z}_{\text{P},n'}^{(J,|\mathcal{A}_\text{tx}|)\ist\text{T}}]\transp$.
The global \ac{lhf} is
\begin{align}
	&f(\V{z}_{\text{A},{1:n}},\V{z}_{\text{P},{1:n}}|\V{y}_{1:n},\V{a}_{\text{A},{1:n}},\V{a}_{\text{P},{1:n}}) \nn\\
	&\hspace{13mm}= \prod_{n'=1}^{n} f(\V{z}_{\text{A},n'}|\V{y}_{n'},\V{a}_{\text{A},n'}) \ist f(\V{z}_{\text{P},n'}|\V{y}_{n'},\V{a}_{\text{P},n'}) \nn\\
	&\hspace{13mm}= \prod_{n'=1}^{n} \prod_{j=1}^{J} \prod_{l=1}^{M_{\text{A},n'}^{(j)}} f_{\text{A}}(\V{z}_{\text{A},n',l}^{(j)}|\V{y}_{n'}, a_{\text{A},n',l}^{(j)}) \nn\\
	&\hspace{17mm} \times \prod_{j' \in \mathcal{A}_{\text{tx}}} \prod_{l=1}^{M_{\text{P},n'}^{(j,j')}} f_{\text{P}}(\V{z}_{\text{P},n',l}^{(j,j')}|\V{y}_{n'}, a_{\text{P},n',l}^{(j,j')})\ist.
	\label{equ:jointLHF_timen}
\end{align}
Using the active measurement \ac{lhf} from Section~\ref{sec:measLikelihood_active} and the passive measurement \ac{lhf} from Section~\ref{sec:measLikelihood_passive}, the association variables $a_{\text{A},n,l}^{(j)}$ and $a_{\text{P},n,l}^{(j,j')}$, the false positive \acp{pdf}, the individual factors in \eqref{equ:jointLHF_timen} are given by
\begin{align}
	{f}_{\text{A}}(\V{z}_{\text{A},n,l}^{(j)}|\V{y}_n, a_{\text{A},n,l}^{(j)})=
	 \begin{cases}
		\frac{f_\text{A}(\V{z}_{\text{A},n,l}^{(j)}|\V{x}_n,\V{X}_n)}{f_{\text{fp}}(\V{z}_{\text{A},n,l}^{(j)})}, &\text{$a_{\text{A},n,l}^{(j)} = 1$}\\ 
		1, & \text{$a_{\text{A},n,l}^{(j)} = 0$}
	\end{cases}
	\label{equ:LHFActive}\\
	{f}_{\text{P}}(\V{z}_{\text{P},n,l}^{(j,j')}|\V{y}_n, a_{\text{P},n,l}^{(j,j')})=
	\begin{cases}
		\frac{f_\text{P}(\V{z}_{\text{P},n,l}^{(j,j')}|\V{x}_n,\V{X}_n)}{f_{\text{fp}}(\V{z}_{\text{P},n,l}^{(j,j')})}, &\text{$a_{\text{P},n,l}^{(j,j')} = 1$}\\ 
		1, & \text{$a_{\text{P},n,l}^{(j,j')} = 0$}
	\end{cases}\ist.
	\label{equ:LHFPassive}
\end{align}
 

\section{Sum-Product Algorithm} \label{sec:algorithm}

\subsection{Problem Statement}

The objective is to track the extended object by estimating the extended object state $\V{y}_n$ in a sequential manner from the observed measurement vectors $\V{z}_{\text{A},1:n}$ and $\V{z}_{\text{P},1:n}$. 
Our Bayesian method aims to calculate the marginal posterior \ac{pdf} $f(\V{y}_{n} | \V{z}_{\text{A},1:n},\V{z}_{\text{P},1:n})$.
Specifically, estimates of the augmented extended object state $\V{y}_n$ are obtained by the \ac{mmse} estimator \cite{Kay1993} as:
\begin{equation}\label{eq:mmse}
	\hat{\V{y}}^\text{MMSE}_{n} \,\triangleq \int \rmv \V{y}_{n} \, f(\V{y}_{n} | \V{z}_{\text{A},1:n},\V{z}_{\text{P},1:n} )\, \mathrm{d}\V{y}_{n}  \,
\end{equation}
with $\hat{\V{y}}^{\text{MMSE}}_n = [\hat{\V{x}}^{\text{MMSE T}}_n  \,\hat{\V{X}}^{\text{MMSE T}}_n]^\text{T}$. 
\subsection{Joint Posterior and Factor Graph}

\begin{figure}[t]
	\centering
	\vspace{-3mm}
	\includegraphics{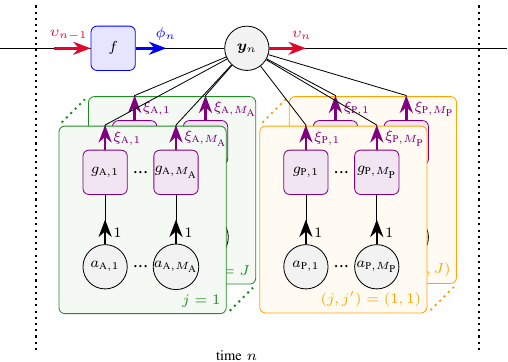}
	\caption{Factor graph representing the factorization of the joint posterior \ac{pdf} in \eqref{equ: joint posterior} and the messages according to the SPA.
	For following short notations are used: $M_{\text{A}} \triangleq M_{\text{A},n}^{(j)}$, $M_{\text{P}} \triangleq M_{\text{P},n}^{(j,j')}$, $a_{\text{A},l} \triangleq a_{\text{A},n,l}^{(j)}$, $a_{\text{P},l} \triangleq a_{\text{P},n,l}^{(j,j')}$, ${g}_{\text{A},l} \triangleq 	{g}_{\text{A}}(\V{z}_{\text{A},n,l}^{(j)}|\V{y}_n, a_{\text{A},n,l}^{(j)})$, ${g}_{\text{P},l} \triangleq 	{g}_{\text{P}}(\V{z}_{\text{P},n,l}^{(j,j')}|\V{y}_n, a_{\text{P},n,l}^{(j,j')}) $, $\xi_{\text{A},l}\triangleq \xi_{\text{A},n,l}^{(j)}(\V{y}_n)$, $\xi_{\text{P},l}\triangleq \xi_{\text{P},n,l}^{(j,j')}(\V{y}_n)$, $f\triangleq f(\V{y}_n|\V{y}_{n-1})$, $\upsilon_n \triangleq \upsilon(\V{y}_n)$, $\phi_n\triangleq \phi(\V{y}_n)$.}
	\label{fig:factorGraph}
	\vspace{-6mm}
\end{figure}

According to Bayes' rule, the introduced independence assumptions, the state transition \ac{pdf} in \eqref{equ:state_transition_model}, the data association factors in \eqref{equ:PMFassocVarA} and \eqref{equ:PMFassocVarP}, and the global \acp{lhf} in \eqref{equ:jointLHF_timen}, the joint posterior \ac{pdf} of $\V{y}_{0:n}$, $\V{a}_{\text{A},1:n}$, and $\V{a}_{\text{P},1:n}$ given the observed (and thus fixed) measurements $\V{z}_{\text{A},1:n}$ and $\V{z}_{\text{P},1:n}$ is given by
\vspace*{-1mm}
\begin{align}
	&f(\V{y}_{0:n},\V{a}_{\text{A},{1:n}}, \V{a}_{\text{P},{1:n}} | \V{z}_{\text{A},{1:n}}, \V{z}_{\text{P},{1:n}})\nn\\ 
	&\propto f(\V{z}_{\text{A},{1:n}},\V{z}_{\text{P},{1:n}}|\V{y}_{1:n},\V{a}_{\text{A},{1:n}},\V{a}_{\text{P},{1:n}}) f(\V{y}_{0:n}, \V{a}_{\text{A},{1:n}}, \V{a}_{\text{P},{1:n}}) \nn\\
	&=f(\V{z}_{\text{A},{1:n}}|\V{y}_{1:n},\V{a}_{\text{A},{1:n}}) f(\V{z}_{\text{P},{1:n}}|\V{y}_{1:n},\V{a}_{\text{P},{1:n}}) \nn\\
	&\hspace{6mm} \times f(\V{y}_{0:n})\eta_{\text{A}}(\V{a}_{\text{A},{1:n}})\eta_{\text{P}}(\V{a}_{\text{P},{1:n}}) \nn\\
	&\propto f(\V{y}_0) \prod_{n'=1}^{n} f(\V{y}_{n'}|\V{y}_{n'-1}) \prod_{j=1}^{J} \prod_{l=1}^{M_{\text{A},n'}^{(j)}} {g}_{\text{A}}(\V{z}_{\text{A},n',l}^{(j)}|\V{y}_{n'}, a_{\text{A},n',l}^{(j)})\nn\\
	&\hspace{6mm} \times \prod_{j' \in \mathcal{A}_{\text{tx}}} \prod_{l=1}^{M_{\text{P},n'}^{(j,j')}} {g}_{\text{P}}(\V{z}_{\text{P},n',l}^{(j,j')}|\V{y}_{n'}, a_{\text{P},n',l}^{(j,j')})
	\label{equ: joint posterior}\\[-7mm]\nn
\end{align}
where $f(\V{y}_n|\V{y}_{n-1})$ is the state transition function.
Fig.~\ref{fig:factorGraph} is the factor graph that represents the factorization of the joint posterior in (\ref{equ: joint posterior}).
The pseudo-likelihood function for the active and the passive case can be represented as
\begin{align}
	{g}_{\text{A}}(\V{z}_{\text{A},n,l}^{(j)}|\V{y}_n, a_{\text{A},n,l}^{(j)}) &\rmv\rmv=\rmv\rmv {f}_{\text{A}}(\V{z}_{\text{A},n,l}^{(j)}|\V{y}_n, a_{\text{A},n,l}^{(j)}) \iist \eta_\text{A}(a_{\text{A},n,l}^{(j)}) \nn\\[0.5mm]
	&\rmv\rmv=\rmv\rmv \begin{cases}
		\frac{\mu_{\text{m}} f_\text{A}(\V{z}_{\text{A},n,l}^{(j)}|\V{x}_n,  \V{X}_n)}{\mu_{\text{fp}} f_{\text{fp}}(\V{z}_{\text{A},n,l}^{(j)})}, \hspace*{-2mm}&\text{$a_{\text{A},n,l}^{(j)} = 1$}\\ 
		1, \hspace*{-2mm}& \text{$a_{\text{A},n,l}^{(j)} = 0$}
	\end{cases}
	\label{equ: active pseudo-measurement likelihood}\\[2mm]
	{g}_{\text{P}}(\V{z}_{\text{P},n,l}^{(j,j')}|\V{y}_n, a_{\text{P},n,l}^{(j,j')}) 
	&\rmv\rmv=\rmv\rmv{f}_{\text{P}}(\V{z}_{\text{P},n,l}^{(j,j')}|\V{y}_n, a_{\text{P},n,l}^{(j,j')})\iist \eta_\text{P}(a_{\text{P},n,l}^{(j,j')}) \nn\\[0.5mm]
	&\rmv\rmv=\rmv\rmv \begin{cases}
		\frac{\mu_{\text{m}} f_\text{P}(\V{z}_{\text{P},n,l}^{(j,j')}|\V{x}_n,  \V{X}_n)}{\mu_{\text{fp}} f_{\text{fp}}(\V{z}_{\text{P},n,l}^{(j,j')})}, \hspace*{-2mm}&\text{$a_{\text{P},n,l}^{(j,j')} = 1$}\\ 
		1, \hspace*{-2mm}& \text{$a_{\text{P},n,l}^{(j,j')} = 0$}
	\end{cases}
	\label{equ: passive pseudo-measurement likelihood}
\end{align}
{The pseudo-likelihood functions ${g}_{\text{A}}(\cdot)$ and ${g}_{\text{P}}(\cdot)$
represent scaled likelihood ratios between the EO-related measurement model and the false positive model, thereby quantifying the evidence that a measurement originates from the object.
Consequently, if a measurement is likely to be EO-related, the corresponding pseudo-likelihood value becomes significantly larger than one when $a_{\text{A},n,l}^{(j)} = 1$ or $a_{\text{P},n,l}^{(j,j')} = 1$,
otherwise, it equals one when the measurement is assumed to be a false positive.}
The state estimate in (\ref{eq:mmse}) is obtained by calculating the marginal density from the joint posterior \acp{pdf} $f(\V{y}_{0:n} | \V{z}_{\text{A},1:n},\V{z}_{\text{P},1:n})$ by applying the \ac{spa}\cite{KschischangTIT2001}.

\subsection{Sum-Product Algorithm} \label{sec:spa}

The factor graph in Fig.~\ref{fig:factorGraph} contains no circles, therefore we can get the exact values for the marginal posterior after applying the \ac{spa} for each time step. Additionally, we assume that messages are not sent backward since the EO states are conditionally independent given the past measurements. 
{The objective of the SPA is to compute the marginal posterior distribution $f(\V{y}_{n} | \V{z}_{\text{A},1:n},\V{z}_{\text{P},1:n} )$. }

The message passing operations are shown in the following equations.
{First, a prediction message is obtained by propagating the previous belief through the state transition model.}
The prediction message is given by
\begin{equation}
	\phi(\V{y}_n) = \int f(\V{y}_{n}|\V{y}_{n-1}) \upsilon(\V{y}_{n-1}) d\V{y}_{n-1}
	\label{equ: prediction}
\end{equation}
where $v(\V{y}_{n-1})$ is the belief of the augmented \ac{eo} state $\V{y}_{n-1}$ from the time $n-1$. After the prediction step, the following calculations are performed for all acnhors $j \in \{1,\dots,J\}$ and all anchor pairs $(j,j')$ with $j \in \{1,\dots,J\}$ and $t \in \mathcal{A}_\text{tx}$ in parallel.

{For every detected multipath component, a soft association message is formed by marginalizing over the binary association variable $a_{\text{A},n,l}^{(j)}$ and $a_{\text{P},n,l}^{(j,j')}$ that indicates whether the measurement is EO-related or a false positive.}
The update messages for the active and passive measurements are given by
\vspace*{-2mm}
\begin{align}
	\xi_{\mathrm{A},n,l}^{(j)}(\V{y}_n) =  \sum_{a_{\text{A},n,l}^{(j)}=0}^{1}{g}_{\text{A}}(\V{z}_{\text{A},n,l}^{(j)}|\V{y}_n, a_{\text{A},n,l}^{(j)}) 
	\label{equ: association_active}\\
	\xi_{\mathrm{P},n,l}^{(j,j')}(\V{y}_n) =  \sum_{a_{\text{P},n,l}^{(j,j')}=0}^{1}{g}_{\text{P}}(\V{z}_{\text{P},n,l}^{(j,j')}|\V{y}_n, a_{\text{P},n,l}^{(j,j')}) 
	\label{equ: association_passive}\\[-8mm]\nn
\end{align}

{Finally, the final belief $\upsilon(\V{y}_n)$ of the augmented \ac{eo} state $\V{y}_n$ is obtained by multiplying the prediction message with all measurement messages.}
\vspace*{-1mm}
\begin{align}
	\upsilon(\V{y}_n) = C_{\text{y},n}\ist\phi(\V{y}_n)\prod_{j=1}^{J} \prod_{l=1}^{M_{\text{A},n}^{(j)}} \xi_{\mathrm{A},n,l}^{(j)}(\V{y}_n)
	\prod_{j' \in \mathcal{A}_{\text{tx}}}\rmv\rmv\prod_{l=1}^{M_{\text{P},n}^{(j,j')}} \xi_{\mathrm{P},n,l}^{(j,j')}(\V{y}_n)
	\label{equ: update}\\[-8mm]\nn
\end{align}
where 
\begin{align}
	\hspace*{-1mm}C_{\text{y},n} \rmv\rmv=\rmv\rmv \left(\int \rmv\rmv\phi(\V{y}_n)\prod_{j=1}^{J} \rmv\rmv\prod_{l=1}^{M_{\text{A},n}^{(j)}} \rmv\rmv\rmv\xi_{\mathrm{A},n,l}^{(j)}(\V{y}_n)\rmv\rmv\rmv
\hspace{-1mm}\prod_{j' \in \mathcal{A}_{\text{tx}}}\rmv\rmv\rmv\rmv\hspace{-1mm}\prod_{l=1}^{M_{\text{P},n}^{(j,j')}}\hspace{-1mm}\rmv\rmv\rmv\xi_{\mathrm{P},n,l}^{(j,j')}(\V{y}_n) \mathrm{d}\V{y}_n \right)^{-1}\rmv\rmv.\\[-8mm]\nn
\end{align}
normalizes the density.
\subsection{Particle Implementation} \label{sec:particle}

Given the nonlinear measurement model and the augmented extended object states, we implement a 
particle filter using a "stacked state" that comprises the kinematic state and extent state \cite{MeyerJSIPN2016}. The belief $\upsilon(\V{y}_n)$ in (\ref{equ: update}) are represented by $P$ particles and corresponding weights, i.e., $\big\{(\V{y}_n^{[p]}, {w_y}_n^{[p]})\big\}_{p=1}^P$.
Each particle $\V{y}_n^{[p]}$ with corresponding weight ${w_y}_n^{[p]}$ is draw from $f(\V{y}_{n}|\V{y}_{n-1}^{[p]})$. The according weights are calculated by
\vspace*{-2mm}
\begin{align}
	w_{y_n}^{[p]} = \Big(\sum_{p=1}^{P}\bar{w}_{y_n}^{[p]}\Big)^{-1}\bar{w}_{y_n}^{[p]}\\[-7mm]\nn
\end{align}
where
\vspace*{-2mm}
\begin{align}
	&\bar{w}_{y_n}^{[p]} = \prod_{j=1}^{J} \prod_{l=1}^{M_{\text{A},n}^{(j)}} \sum_{a_{\text{A},n,l}^{(j)}=0}^{1}{g}_{\text{A}}(\V{z}_{\text{A},n,l}^{(j)}|\V{y}_{n}^{[p]},a_{\text{A},n,l}^{(j)})\nn\\
	&\hspace{10mm} \times \prod_{j' \in \mathcal{A}_{\text{tx}}} \prod_{l=1}^{M_{\text{P},n}^{(j,j')}} \sum_{a_{\text{P},n,l}^{(j,j')}=0}^{1}{g}_{\text{P}}(\V{z}_{\text{P},n,l}^{(j,j')}|\V{y}_{n}^{[p]}, a_{\text{P},n,l}^{(j,j')})\ist.
	\label{equ: kinematic weights}\\[-7mm]\nn
\end{align}
The approximate \ac{mmse} estimate of $\V{y}_n$ is then given as $\hat{\V{y}}^\text{MMSE}_{n} \approx \sum_{p=1}^P \RV{y}_n^{[p]} w_{y_n}^{[p]}$.  To avoid particle degeneracy, a resampling
step \cite{ArulampalamTSP2002} is performed at each time $n$.


\section{Results}\label{sec:results}

\begin{figure}[t]	
	\centering
	\setlength{\abovecaptionskip}{0pt}
	\setlength{\belowcaptionskip}{0pt}
	\vspace{-6mm}
    \includegraphics{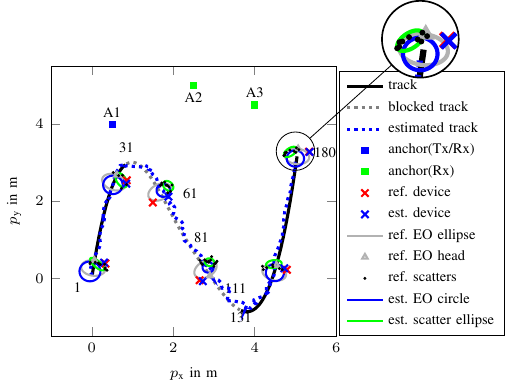}
	\caption{Graphical representation of the synthetic trajectory and an example realization obtained with the AP-PROP-SMP method. The scatter points are generated with respect to anchor pair (A1, A3) for passive measurements.
	}\label{fig:simulationScenario}
	\vspace{-3mm}
\end{figure}
The performance of the proposed algorithm is evaluated  using both synthetic measurements and real radio signals. The synthetic data are obtained from numerical simulation, and the \ac{mmse} estimate of each state variable  is compared with the true value in the setup. 
\begin{figure*}[t]
	\captionsetup[subfloat]{captionskip=-2mm}
	\centering
	\setlength{\abovecaptionskip}{0pt}
	\setlength{\belowcaptionskip}{0pt}
	\includegraphics{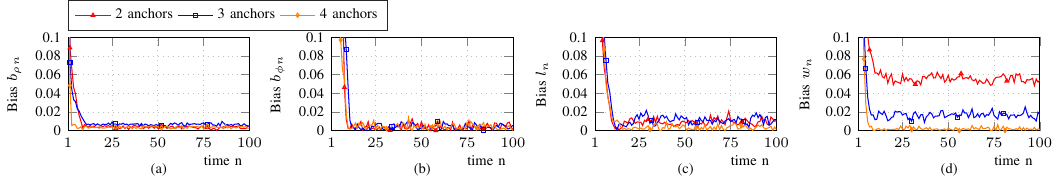}
	\caption{Absolute bias of state variables $\bm{b}_n$ and $\bm{X}_n$  with varying number of anchors over numerical simulations.} \label{fig:ext_static}
	\vspace{-3mm}
\end{figure*}
\subsection{Common Setup Analysis} \label{sec:setup}

The following setup and parameters are commonly used for all analyses presented.
The number of particles is set to $P = 5000$, and the particles consist of all considered random variables. 
The state noise parameters are set to ${\sigma_{\text{a}}} = 2\,\mathrm{m/s^2}$, $\kappa_{\rho}=100$, $\sigma_{\phi} = 0.5\,\mathrm{rad}$, $\kappa_{\text{l}} = \kappa_{\text{w}} =\kappa_{\delta} = \kappa_{\text{r}} = \kappa_{\text{w}_\text{s}} = 400$. 
{The parameter $\kappa$ controls the convergence behavior of the extent states. A larger $\kappa$ reduces the deviation from the previous state estimate, resulting in a more conservative update. Consequently, if the current estimate is far from the true value, a larger $\kappa$ leads to slower convergence.
The effectiveness of the selected values of $\kappa$ is demonstrated by the results shown in Fig.~\ref{fig:ext_static}.}
Besides, $P_{\text{mix}}=0.5$.

The results are shown in terms of the \ac{rmse} and \ac{ecdf} over $100$ simulation runs.
The \ac{rmse} of the estimated device position is calculated by $e_{n}^{\text{RMSE}}~=~\sqrt{\E{\norm{\hat{\bm{m}}^{\text{MMSE}}_n -\bm{m}_n}{2}}}$.
The \ac{rmse} values are averaged over all $100$ realizations, while the \ac{ecdf} of the \ac{rmse} is also averaged over all time steps. 
The \ac{crlb} is calculated as a performance baseline on the position error variance considering all direct LOS links in active measurements of all $J$ anchors at a single time step $n$\cite{WitrisalJWCOML2016}. 
We provide the corresponding \ac{pcrlb} that additionally considering the information provided by the state transition model of the agent state $\V{x}_n$\cite{Tichavsky1998}.
{Detailed derivation of the \ac{pcrlb} can be found in \cite[Eqs. (32)-(33)]{Venus2024}}.
{Specifically, the P-CRLB* is calculated by assuming direct LOS paths that are always available at all time $n$ in active measurements of all anchors, regardless of the actual OLOS periods along the trajectory.  It quantifies the achievable accuracy if LOS information were never lost.
In contrast, P-CRLB accounts for the actual visibility of LOS paths, i.e., the LOS availability varies according to the blockage conditions as described in Sec.~\ref{sec:synthetic_setup} and Sec.~\ref{sec:evaluation_real_meas}. During OLOS, the Fisher information contributed by direct paths is removed, so the bound becomes looser.}
{Note that the \ac{pcrlb} accounts for  \ac{los} measurements only. 
In practice, the particle-based estimators exploit additional information from active and passive scattering components that are not captured in the Fisher information calculation. As a result, the estimator \ac{rmse}  occasionally drops slightly below the bound, which does not violate the \ac{pcrlb} principle but reflects a modeling mismatch between the bound and the estimator.}

\subsection{Reference Methods}\label{sec:refMethods}

In the following sections, we compare the proposed multi-sensor, multiple-measurement \ac{pda} algorithm (PROP) in three different settings, referred to as {A-PROP-SMP, AP-PROP-SMP} and AP-PROP. The prefix ``A-'' indicates that only active measurements are used, whereas ``AP-'' denotes the use of both active and passive measurements. The suffix ``-SMP'' indicates the use of the {simplified \ac{eo} model}, while the absence of the suffix indicates the use of the \ac{eo} model.
{We compare the PROP algorithm with a EO-based algorithm that uses a traditional elliptical EO model \cite{Koch2008, MeyerTSP2021,Zhu2024}, which assumes that scatter points are uniformly distributed over the entire object volume, independent of the anchor geometry.
We refer to these variants as A-TRAD and AP-TRAD in Section~\ref{sec: results_synthetic_fully}.
The likelihood functions in the traditional model integrates over the full ellipse support, whereas the proposed EO model integrates only over an anchor-dependent visible sector.}
We also consider the classical \ac{pda} algorithm\cite{BarShalomTCS2009} using the point-object assumption-based \acp{lhf} (see \eqref{equ:measLikelihood_active_point} and \eqref{equ:measLikelihood_passive_point} in Appendix~\ref{sec:PDA_algorithm}) {as benchmark methods}. 
They are evaluated with {our particle implementation} in three settings: A-PDA and AP-PDA with $\sigma_{\text{r}} = 0$, and AP-PDA2 with $\sigma_{\text{r}} = 0.2,\mathrm{m}$, 
The variance $\sigma_{\text{r}}$ captures the variance caused by the \ac{eo}. 
The value of $\sigma_{\text{r}}$ in A-PDA and AP-PDA is chosen to be 0, which means the conventional \ac{pda} method is applied without considering the affect of the \ac{eo}, while the value of $\sigma_{\text{r}}$ in AP-PDA2 is chosen to approximately reflect the size of a human torso.
The number of samples for importance sampling in (\ref{equ:measLikelihood_idealAS_appr}) and (\ref{equ:measLikelihood_idealPS_appr}) is chosen to be $I=100$ as default.

\subsection{Evaluation with Synthetic Measurements}
\label{sec:synthetic_setup}
The algorithm is evaluated in the scenario presented in Fig.~\ref{fig:simulationScenario}. The \ac{eo} is modeled as an ellipse with the semi-major axis $l = 0.3\,\mathrm{m}$ and the semi-minor axis $w = 0.2\,\mathrm{m}$. It moves along a smooth trajectory with two direction changes.
The radio device is coupled with the EO with fixed distance $b_{\rho} = 0.32\,\mathrm{m}$ and direction $b_{\phi} = -\pi/3$ with respect to the EO coordinate.
Scatter points are uniformly distributed within the elliptical annular sector $ \mathcal{S}^{(j)}(\V{X}_n,\V{x}_n) $ and $  \mathcal{S}^{(j,j')}(\V{X}_n,\V{x}_n) $ for active measurements and passive measurements, respectively, as described in Appendix \ref{scatterPDF}.
The mean number of object-related measurements (scattering measurements) is set as a constant $\mu_{\text{m}} = 5$ considering the limited size of the EO, and the scatter points are regenerated independently for each time step and each anchor, for active and passive measurements, respectively. The mean number of false positive measurements is chosen to be $\mu_{\text{fp}}=5$.

The object is observed at $180$ discrete time steps 
at a constant observation interval of $\Delta \text{T} = 100\,\mathrm{ms}$.
For active measurements, the signals are transmitted from the radio device and  received at $3$ anchors. For passive measurements, the signals are transmitted from anchor A1  and received at all anchors A1, A2, and A3.
We assume that active measurements are not continuously available due to the blockage of the EO when it is moving on the trajectory. 
More specifically, active measurements of A1 are missing during time steps $[31, 80]$ and $[111,130]$, those of A2  are missing during $[31,130]$, and those of A3 anchor are missing during $[31, 60]$ and $[111,130]$.  
For the passive case, all observations are obtained from scatter points, and they are consistently available along the whole trajectory. 

\subsubsection{Studies on the Extent State Estimation}\label{sec: extStates_static}
{We first investigate the estimation of the bias and extent states using the EO model (see Sec.~\ref{sec:ideal_concept}) in a static scenario with both active and passive measurements. The EO center is fixed at $(0,0)$ for all time steps $n$. To study the influence of anchor geometry, we consider three configurations:
(1) two anchors located at $(1.5,1)$ and $(-1.5,1)$;
(2) three anchors located at $(1.5,1)$, $(0,2)$, and $(-1.5,1)$;
(3) four anchors located at $(2,0)$, $(0,2)$, $(-2,0)$, and $(0,-2)$.
The resulting absolute bias\footnote{Mathematically, the absolute bias is defined as $\text{bias} = \left| \mathbb{E}[\hat{\theta}] - \theta \right|$, where $\hat{\theta}$ denotes the estimator and $\theta$ the true parameter value.} with respect to the reference values in Sec.~\ref{sec:synthetic_setup}, obtained from 100 simulation runs over 100 time steps, is shown in Fig.~\ref{fig:ext_static}. 
As the number of anchors increases, the spatial diversity of viewing angles improves the geometric observability of the EO extent, reducing parameter ambiguity. In particular, the four-anchor configuration provides sufficient angular coverage to constrain both axes of the object, leading to consistent convergence of the estimated EO size to the ground-truth values.}

\subsubsection{Fully Synthetic Measurements}\label{sec: results_synthetic_fully}

\begin{figure}
	\captionsetup[subfloat]{captionskip=-2mm}
	\centering
	\setlength{\abovecaptionskip}{0pt}
	\setlength{\belowcaptionskip}{0pt}
	\subfloat[\label{fig:rmse_noCE_500}]{\includegraphics{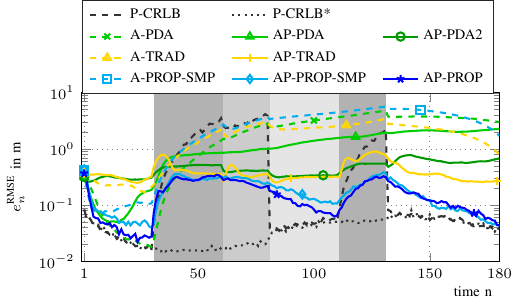}}\vspace{-0mm}
	\subfloat[\label{fig:cdf_noCE_500}]{\includegraphics{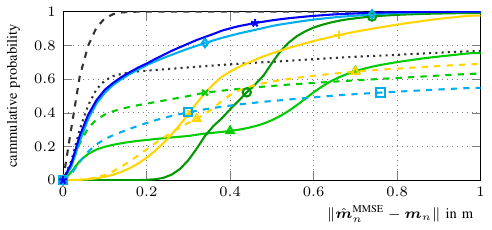}}\vspace{-0mm}
	\subfloat[\label{fig:ext_noCE_500}]{\includegraphics{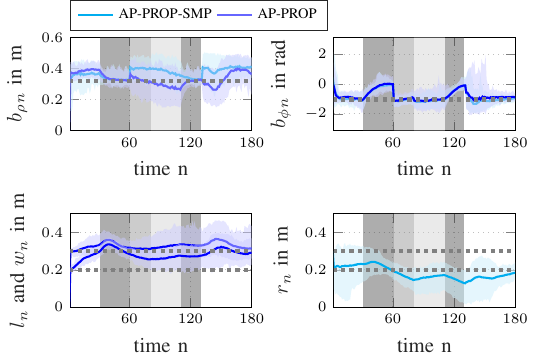}}
	\caption{Results of different methods in fully synthetic measurements with $500\,\mathrm{MHz}$ bandwidth, (a) is the \ac{rmse} of the estimated position $\bm{m}_n$, (b) is the \ac{ecdf} of \ac{rmse} based on numerical simulations, and {(c) is estimation performance of extent states varying with time $n$ of AP-PROP-SMP and AP-PROP. The solid lines depict the empirical mean over 100 simulations, and the shaded regions indicate the 95 percentile-based confidence intervals.} Gray shades indicate the number of anchors missing active measurements, as described in \ref{sec:synthetic_setup}.} \label{fig:syn_noCE_500}
	\vspace{-3mm}
\end{figure}

\begin{figure*}[t]
	\captionsetup[subfloat]{captionskip=-2mm}
	\centering
	\setlength{\abovecaptionskip}{0pt}
	\setlength{\belowcaptionskip}{0pt}
	\subfloat[\label{fig:active}]{\includegraphics{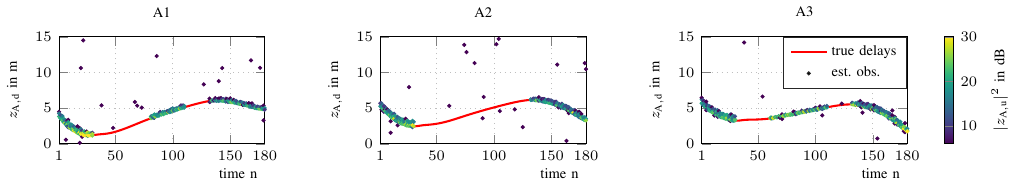}}\vspace{-0mm}
	\subfloat[\label{fig:passive}]{\includegraphics{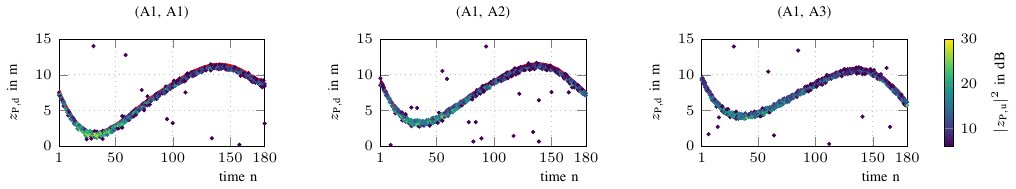}}
	\caption{Measurements (estimated delays and amplitudes from CEDA) of one realization along the whole trajectory with $500\,\mathrm{MHz}$ bandwidth. (a) shows active observations from anchor A1, A2, and A3, respectively, and (b) shows passive observations from anchor pair (A1, A1), (A1, A2), and (A1, A3), respectively. The red line indicates the true delays used for generating the radio signals. 
	}\label{fig:observations}
	\vspace{-6mm}
\end{figure*}
\begin{figure*}[t]
	\captionsetup[subfloat]{captionskip=-2mm}
	\centering
	\setlength{\abovecaptionskip}{0pt}
	\setlength{\belowcaptionskip}{0pt}
	\subfloat[\label{fig:rmse_newCE_500}]{\includegraphics{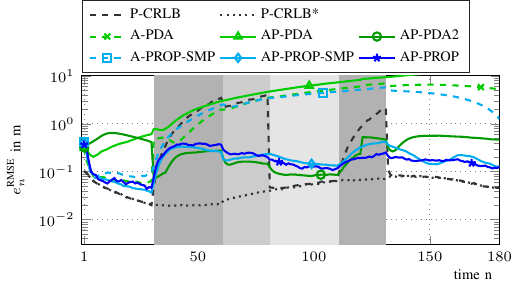}}
	\subfloat[\label{fig:cdf_newCE_500}]{\includegraphics{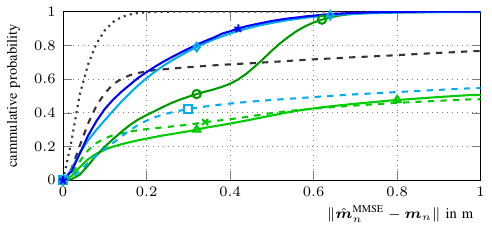}}\vspace{-3mm}
	\subfloat[\label{fig:rmse_newCE_3G}]{\includegraphics{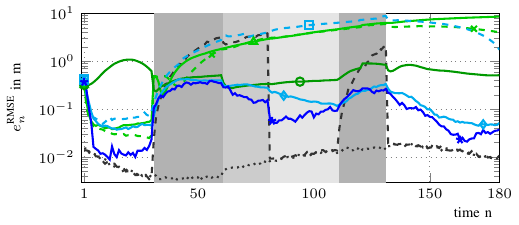}}
	\subfloat[\label{fig:cdf_newCE_3G}]{\includegraphics{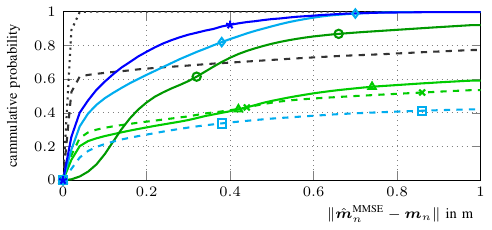}}
	\caption{Numerical results of different methods in synthetic radio measurements with $500\,\mathrm{MHz}$ bandwidth in (a),(b), and $3\,\mathrm{GHz}$ in (c), (d). While (a), (c) are the \ac{rmse} of the estimated device position $\bm{m}_n$, and (b), (d) are the \ac{ecdf} of \ac{rmse} based on numerical simulations. Gray shades indicate the number of anchors missing active measurements, as described in \ref{sec:synthetic_setup}.} \label{fig:syn_newCE}
	\vspace{-3mm}
\end{figure*}
In this experiment, measurements are drawn directly from the statistical measurement model (i.e., not extracted from the raw radio signals). For each simulation, both LOS- and scattering-related delay measurements are generated using the geometry and \ac{eo} model parameters, and measurement noise is added with a variance determined by the Fisher information and the normalized-amplitude measurements described in Section~\ref{sec:measurement_model}. Both LOS- and scattering-related complex amplitudes are generated from the geometry and \ac{eo} model parameters and decay according to free-space path loss, cf.~\eqref{equ:amp_active_LOS}, \eqref{equ:amp_active_scatter}, and \eqref{equ:amp_passive_scatter}. The nominal magnitudes are set to $|\alpha_{\text{A}}|=1$ and $|\alpha_{\text{P}}|=1$, and the scattering coefficient is $\beta=0.5$. The \ac{snr} at $1\,\mathrm{m}$ from the radio device is set to $30\,\mathrm{dB}$, yielding a radio-signal noise variance of $10^{-3}$, and the carrier frequency is $f_c=6.95\,\mathrm{GHz}$. 
The mean values of the normalized amplitudes (defined as the square root of the component \ac{snr}) are determined by the magnitudes of the complex amplitudes together with the noise variance of the radio signal~\cite{WitrisalJWCOML2016,LiTWC2022}. 
The normalized-amplitude measurements are generated using a Rician \ac{pdf} with squared scale parameter $1/2$ and the mean values of the normalized amplitudes ~\cite[Ch.~3.8]{Kay1993} \cite{LiTWC2022}. The detection threshold is $\gamma=2$ ($6\,\mathrm{dB}$), and the bandwidth is $500\,\mathrm{MHz}$.

The performance of different methods is shown in Fig.~\ref{fig:syn_noCE_500}\footnote{{For the active-only case, performance is limited by missing LOS measurements during OLOS periods, making A-PROP and A-PROP-SMP nearly identical; thus, only A-PROP-SMP is shown.} }.
Fig.~\ref{fig:rmse_noCE_500} shows the \ac{rmse} as a function of the discrete observation time $n$, with light gray indicating missing active measurements from one anchor, medium gray indicating those from 2 anchors, and dark gray indicating those from all 3 anchors, corresponding to the description in Sec.~\ref{sec:synthetic_setup}.
Fig.~\ref{fig:cdf_noCE_500} is the \acp{ecdf} of the \ac{rmse}.
It can be observed from the figures that AP-PROP closely attains the P-CRLB before the OLOS situation and converges back to the P-CRLB afterwards.
The performance of the joint estimation AP-PROP-SMP significantly outperforms the active-only estimation A-PROP-SMP during and after the blocked time steps. 
In comparison, AP-PROP-SMP achieves a performance comparable to AP-PROP, despite employing a simplified \ac{eo} model.
{Moreover, with the traditional elliptical EO inference model\cite{Koch2008, MeyerTSP2021,Zhu2024}, AP-TRAD exhibits consistently higher estimation errors over the entire trajectory. The traditional model implicitly assumes isotropic scatter distribution, which overestimates the measurement support, leading to biased estimation.}
{With the inflated likelihood variance in  \eqref{equ:measLikelihood_active_point} and \eqref{equ:measLikelihood_passive_point} in Appendix~\ref{sec:PDA_algorithm}, AP-PDA2 shows reduced errors compared with AP-PDA, during and after the blocked time steps.}
{Note that A-PDA may outperform A-PROP-SMP in the initial phase of the trajectory because EO modeling offers no advantage when direct LOS paths are available. These active-only methods diverge in the \ac{olos} region due to the absence of informative active measurements, and they barely recover thereafter.}
{Fig.~\ref{fig:ext_noCE_500} presents a comparison of MMSE estimates for state variables of the EO model and the simplified EO model. It shows the empirical mean and confidence interval of the bias and extent states over 100 simulations.
It can be found that both AP-PROP and AP-PROP-SMP demonstrate reliable performance in estimating the bias and extent variables, with reasonably small biases during the OLOS situation.}

\subsubsection{Synthetic Radio Signals}\label{sec: results_synthetic_radio}
\begin{figure}[htbp] 
	\centering 
	
	\captionsetup[subfloat]{captionskip=-2mm}
	\setlength{\abovecaptionskip}{0pt}
	\setlength{\belowcaptionskip}{0pt}
	{\includegraphics{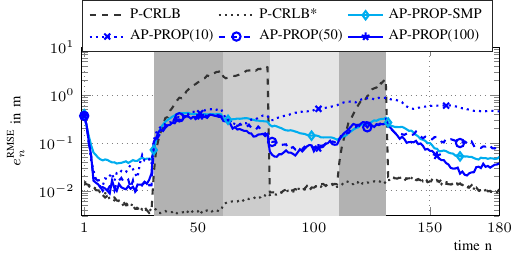}}
	\caption{Comparative \ac{rmse} performance illustrates the influence of different number of samples $I ={10,50,100}$ used in AP-PROP, compared to AP-PROP-SMP with $3\,\mathrm{GHz}$ bandwidth in synthetic radio measurements.}
	\label{fig:syn_newCE_ideal}
	
	\vspace{2mm} 
	
		\captionsetup{type=table} 
		\renewcommand{\baselinestretch}{1}\small\normalsize
		\setlength{\tabcolsep}{3pt} 
		\renewcommand{\arraystretch}{1} 
		\footnotesize
		\captionof{table}{Average runtime per time step $n$ and average \ac{rmse} of different methods with $3\,\mathrm{GHz}$ bandwidth in synthetic radio measurements.}
		\label{tbl:execution_times}
		\begin{tabular}{ r c c c c} 
			\toprule
			{Method}  & { \ac{rmse} (m)} & {runtime (s)} \\
			\midrule
			\multicolumn{1}{r}{AP-PROP(10)} &0.428& 0.414 \\
			\multicolumn{1}{r}{AP-PROP(50)} &0.157& 0.630 \\
			\multicolumn{1}{r}{AP-PROP(100)}&0.142& 1.313 \\
			\multicolumn{1}{r}{AP-PROP-SMP} &0.191& 0.352 \\
			\bottomrule
		\end{tabular}
		\vspace{-4mm}
		
	\end{figure}
In this experiment, the measurements $\V{z}_{\text{A},n}^{(j)}$ and $\V{z}_{\text{P},n}^{(j,j')}$ are obtained by applying a snapshot-based \ac{ceda}~\cite{Hansen2018, Grebien2024} to synthetic radio signals with a detection threshold $\gamma=2$ ($6\,\mathrm{dB}$) and a maximum detection distance $d_{\text{max}} = 30\,\mathrm{m}$. The synthetic radio signals follow the models described in Appendix~\ref{sec:RSmodel}. The transmitted complex baseband signal $s(t)$ is root-raised-cosine with roll-off $0.6$, and two bandwidth settings are considered: $500\,\mathrm{MHz}$ and $3\,\mathrm{GHz}$. The complex amplitudes of the LOS and scattered paths for active and passive measurements are generated according to \eqref{equ:amp_active_LOS}, \eqref{equ:amp_active_scatter}, and \eqref{equ:amp_passive_scatter}, respectively, with nominal magnitudes $|\alpha_{\text{A}}|=1$, $|\alpha_{\text{P}}|=1$, and scattering coefficient $\beta=0.5$. The carrier frequency is $f_c=6.95\,\mathrm{GHz}$, and the \ac{snr} at $1\,\mathrm{m}$ from the radio device is set to $30\,\mathrm{dB}$. Representative examples of measurement estimates extracted from the radio signals are shown in Fig.~\ref{fig:observations}.

The performance of different methods and different bandwidths is shown in Fig.~\ref{fig:syn_newCE}. 
{Because the measurements are extracted from simulated signals, the bandwidth directly determines the delay-estimation resolution of the channel estimation algorithm.}
Fig.~\ref{fig:rmse_newCE_500} and Fig.~\ref{fig:cdf_newCE_500} show the \ac{rmse} as a function of the discrete time $n$ and the \ac{ecdf} of the \ac{rmse} with $500\,\mathrm{MHz}$ bandwidth.
The figures indicate that both AP-PROP and AP-PROP-SMP demonstrate promising performance by incorporating active and passive measurements and employing the multiple-measurement PDA. 
{From Fig.~\ref{fig:cdf_newCE_500} we can see that the performance gain of AP-PROP over AP-PROP-SMP remains limited because the available delay measurements are not informative enough to fully exploit the more accurate EO model.}
The performance of A-PDA and AP-PDA is suboptimal, which can be attributed to the limitations of applying the conventional PDA without accounting for the EO effect.
{AP-PDA2 achieves a higher cumulative probability for small estimation errors than A-PDA and AP-PDA, demonstrating a clear improvement in robustness due to the additional variance introduced. Nevertheless, its performance remains inferior to the proposed AP-PROP method.}
Besides, the performance of A-PROP-SMP is also unsatisfactory due to the absence of passive measurements to provide support under \ac{olos} conditions.
Fig.~\ref{fig:rmse_newCE_3G} and Fig.~\ref{fig:cdf_newCE_3G} show the \ac{rmse} as a function of the discrete time $n$ and the \ac{ecdf} of the \ac{rmse} with $3\,\mathrm{GHz}$ bandwidth.
{Comparing the Fig.~\ref{fig:cdf_newCE_500} and Fig.~\ref{fig:cdf_newCE_3G}, we observe that AP-PROP outperforms AP-PROP-SMP more significant at higher bandwidths, because the more accurate EO model used in AP-PROP can better exploit the finer delay information available in the measurements.}
AP-PDA2 does not achieve comparable performance under this condition.
Similarly, A-PDA, A-PROP-SMP, and AP-PDA do not achieve optimal performance due to the lack of passive measurements and the limitations of the conventional PDA.
 
Fig.~\ref{fig:syn_newCE_ideal} and Table~\ref{tbl:execution_times} present a comparative analysis of the average runtime across time $n$, and the \ac{rmse} performance of the AP-PROP-SMP and AP-PROP evaluated with varying sample numbers ($I = {10, 50, 100}$)\footnote{AP-PROP with different sample sizes are denoted as "AP-PROP(10)", "AP-PROP(50)", and "AP-PROP(100)", respectively, in Fig.~\ref{fig:syn_newCE_ideal} and Table~\ref{tbl:execution_times}.} as given in \eqref{equ:measLikelihood_idealAS_appr} and \eqref{equ:measLikelihood_idealPS_appr}, at  $3\,\mathrm{GHz}$. 
The result shows that  increasing the number of samples enhances the performance of AP-PROP. 
For instance, the averaged RMSE of AP-PROP with $I=100$ is about one-third of that with $I=10$.
While improved accuracy is observed with larger sample sizes, this comes at the expense of increased computational cost. Specifically, using $I=100$ results in a runtime approximately four times greater than that of the AP-PROP-SMP, although this increase yields a lower \ac{rmse} across the entire trajectory.

\subsection{Evaluation with Measured Radio Signals}\label{sec:evaluation_real_meas}

\begin{figure}[t]
	\centering
	\setlength{\abovecaptionskip}{0mm}
	\setlength{\belowcaptionskip}{0mm}
	\subfloat[\label{fig:EO}]{\includegraphics[height=0.25\textwidth]{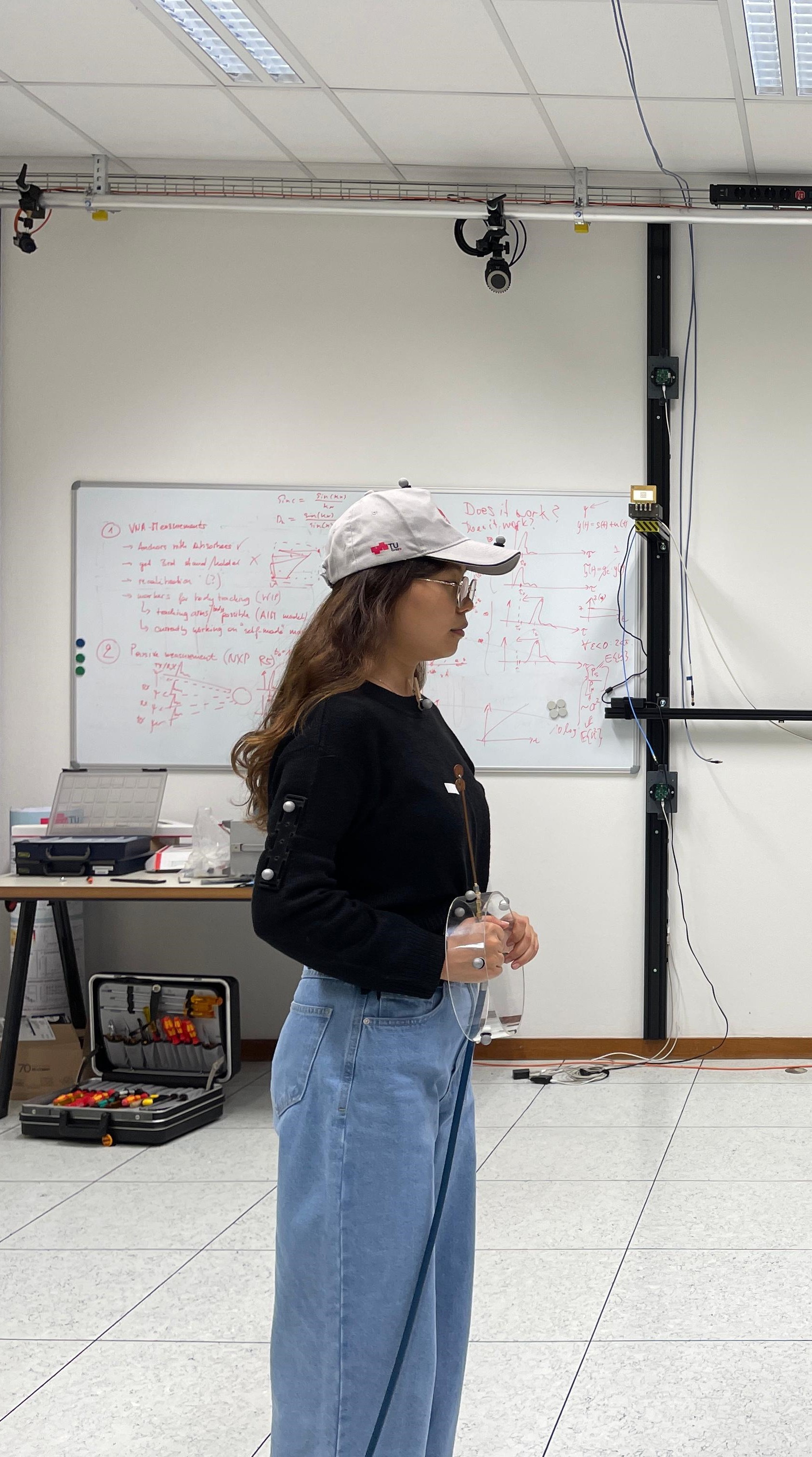}}\vspace{2mm}
	\subfloat[\label{fig:setup}]{\includegraphics[height=0.25\textwidth]{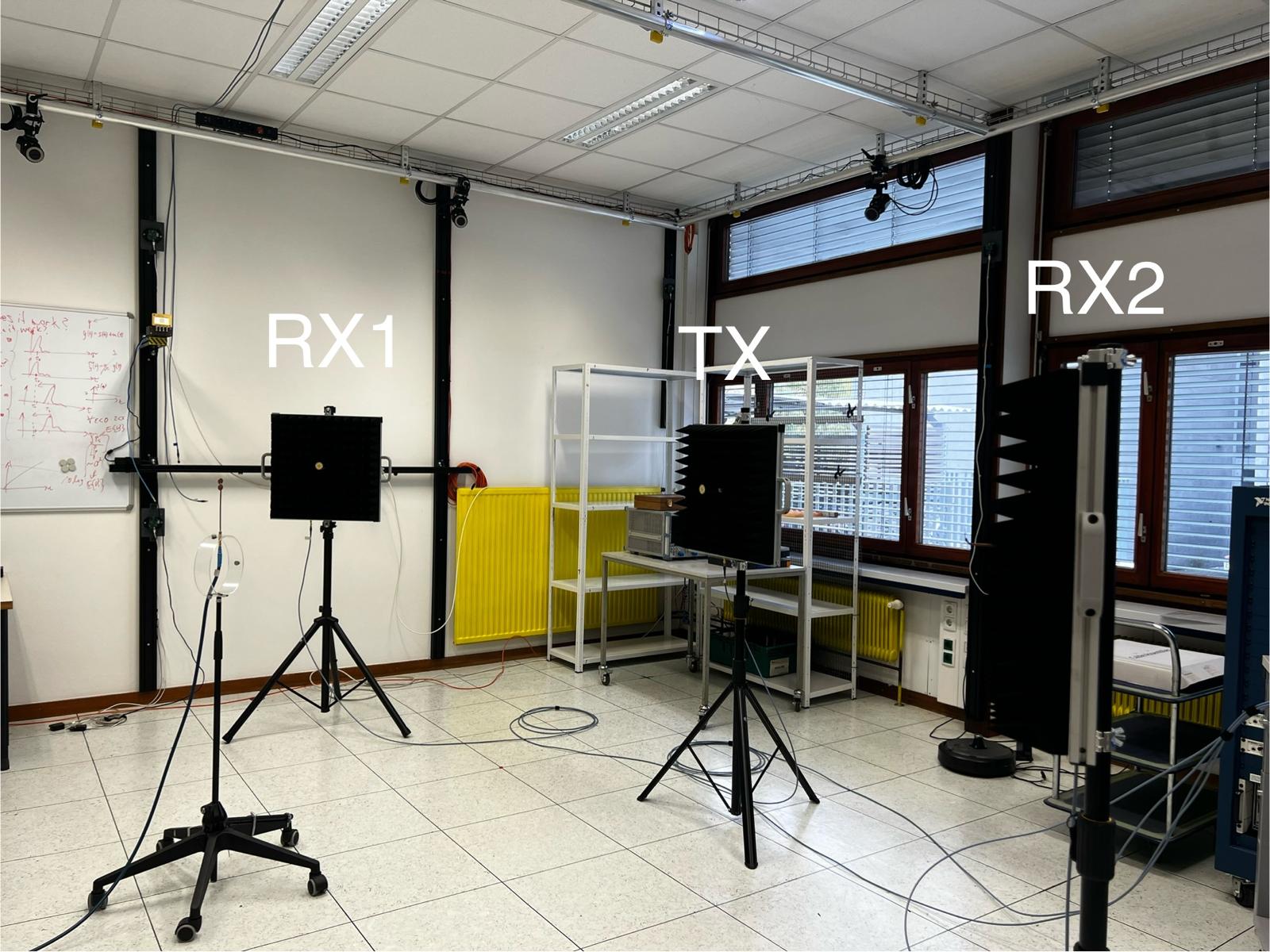}}
	\setlength{\figurewidth}{0.28\textwidth}
	\setlength{\figureheight}{0.28\textwidth}
	\caption{Setup for real radio measurements described in Section~\ref{sec:evaluation_real_meas}. The extended object attended in the measurement campaign is shown in photo (a), and the overall view of the measurement scenario is shown in photo (b).}\label{fig:measurement_setup}
	\vspace{-3mm}
\end{figure}

\begin{figure*}[t]
	\centering
	\setlength{\abovecaptionskip}{0pt}
	\setlength{\belowcaptionskip}{0pt}
	\subfloat[\label{fig:cs_active_3G}]{\includegraphics{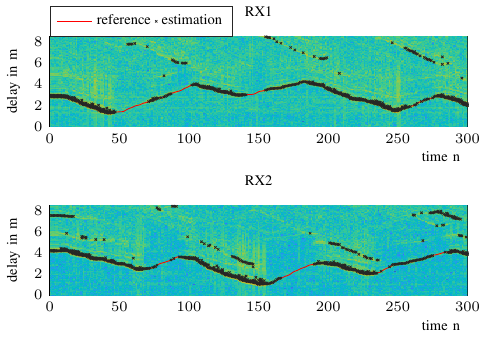}}
	\subfloat[\label{fig:cs_passive_3G}]{\includegraphics{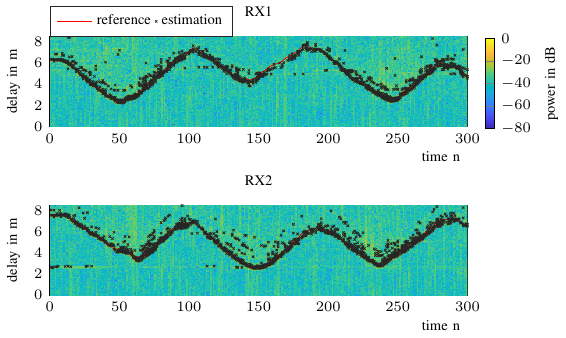}}\vspace{-3mm}
	\subfloat[\label{fig:cs_active_500M}]{\includegraphics{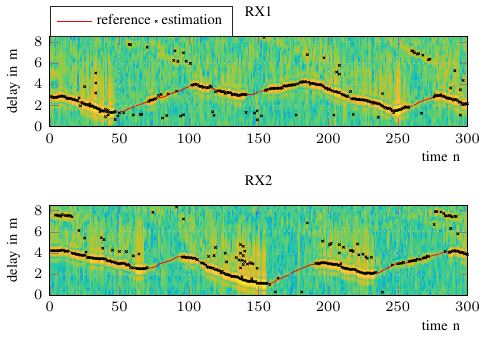}}
	\subfloat[\label{fig:cs_passive_500M}]{\includegraphics{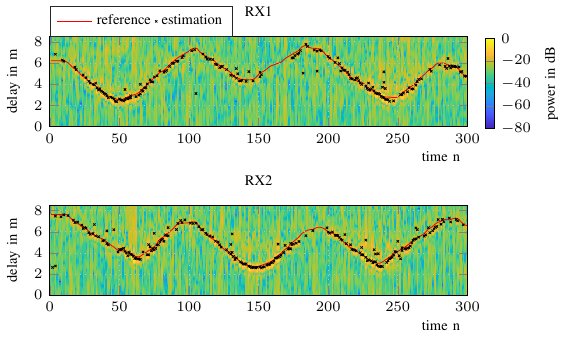}}
	\caption{Observations (estimated delays from CEDA) of real radio measurements as described in Section~\ref{sec:evaluation_real_meas}. Background is the normalized \ac{cir} from each received anchor. (a) and (b) show the active and passive observations with $3\,\mathrm{GHz}$ bandwidth, respectively,  while(c) and (d) show the active and passive observations with $500\,\mathrm{MHz}$ bandwidth, respectively.  The red line indicates the reference delay values obtained from the Qualisys system. {The structured scatter-path delays in the CIR provide empirical support for the assumed surface-scattering distribution of the proposed EO model.}}\label{fig:measObs}
	\vspace{-3mm}
\end{figure*}

\begin{figure*}
	\captionsetup[subfloat]{captionskip=-2mm}
	\centering
	\setlength{\abovecaptionskip}{0pt}
	\setlength{\belowcaptionskip}{0pt}
	\subfloat[\label{fig:rmseMeas3G}]{\includegraphics{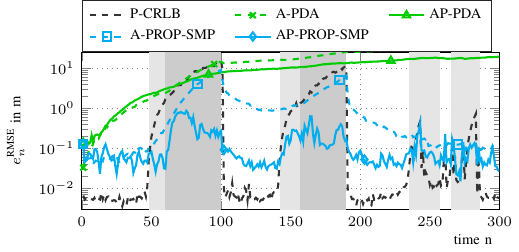}}
	\subfloat[\label{fig:cdfMeas3G}]{\includegraphics{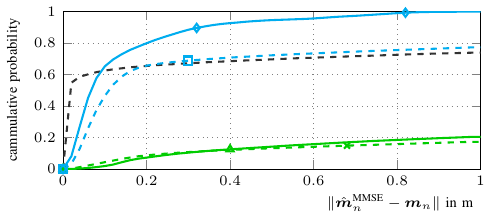}}\vspace{-4mm}
	\subfloat[\label{fig:rmseMeas500M}]{\includegraphics{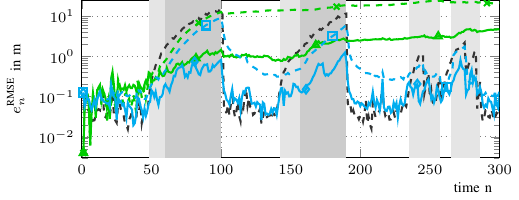}}
	\subfloat[\label{fig:cdfMeas500M}]{\includegraphics{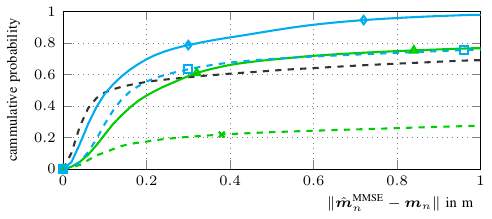}}
	\caption{Numerical results of different methods in real radio measurements with $3\,\mathrm{GHz}$ bandwidth in (a) and (b), and $500\,\mathrm{MHz}$ in (c) and (d). While (a) (c) are the \ac{rmse} of the estimated device position $\bm{m}_n$, and (b) (d) are the \ac{ecdf} of the \ac{rmse} based on numerical simulations. Gray shades indicate the number of anchors missing active measurements, as described in \ref{sec:evaluation_real_meas}.}
    \label{fig:CSmeas}
	\vspace{-3mm}
\end{figure*}

The proposed algorithm is also evaluated with real radio measurements, which have been conducted in the NXP laboratory at TU Graz, Austria. 
As shown in Fig.~\ref{fig:EO}, a person holding a planar coin-shaped dipole antenna near the body is moving along a random trajectory in the room. 
The person is modeled as the \ac{eo}, and the dipole antenna represents the radio device.
Three anchors are deployed, with two serving as receivers (denoted RX1 and RX2) and one as the transmitter for passive measurements (denoted TX). Each anchor is equipped with a XETS antenna\cite{Costa2009}, and an absorber is mounted on the backside of each antenna to suppress backscattering from walls and windows, as shown in Fig.~\ref{fig:setup}.  
The radio signal is generated and captured by an \ac{uwb} M-sequence correlative channel sounder with a frequency range of $3.1-10.6\,\mathrm{GHz}$.
It includes a \ac{uwb} signal generator and two synchronously operating receivers.
A power splitter is used for synchronously transmitting signals for active and passive measurements. 
A long cable is connected to the splitter to transmit signals for active measurements with a delay to seperate the \acp{cir} of the two sets of measurements in time domain. 
The received signal band was selected by a filter with raised-cosine (RC) impulse response $s(t)$, with a roll-off factor of $0.6$, two-sided $3\,\mathrm{dB}$ bandwidths of $3\,\mathrm{GHz}$ and $500\,\mathrm{MHz}$, and a center frequency of $6.95\,\mathrm{GHz}$.
The maximum detection distance $d_{\text{max}} = 30\,\mathrm{m}$.   
The measurement frequency for both active and passive measurements is $5\,\mathrm{Hz}$, leading to an observation rate of $\Delta T=0.2\,\mathrm{s}$. The measurement period is $60\,\mathrm{s}$ in total, which leads to $300$ measurements.
Additionally, a Qualisys motion capture system is used to obtain the reference positions of the human body and the radio device along the whole trajectory. 

The received \acp{cir} with different bandwidths of both active measurements and passive measurements are illustrated in Fig.~\ref{fig:measObs}.  
Estimated delays, obtained after applying the \ac{ceda} \cite{Hansen2018, Grebien2024} on the received signals, are compared with the ground truth values from the Qualisys system for all time $n$.
Figs.~\ref{fig:cs_active_3G} and \ref{fig:cs_passive_3G} show that with $3\,\mathrm{GHz}$ bandwidth, abundant scattering components reflected from the human body are evident in both active and passive measurements. 
The extracted delays of the active and passive \acp{cir} reveal a clear trajectory of the measurement campaign.   
The partial absence of active components indicates the blockage due to the human body between the radio device and anchors. 
As presented in Fig.~\ref{fig:cs_active_500M} and Fig.~\ref{fig:cs_passive_500M}, when the bandwidth decreases to $500\,\mathrm{MHz}$, the scattering components exhibit lower time-delay resolution compared to the $3\,\mathrm{GHz}$ case, resulting in a less detailed scattered \acp{mpc} structure. 
{The extracted \acp{mpc} in Fig.~\ref{fig:measObs} exhibit direction-dependent delay patterns that vary with anchor geometry and remain confined to narrow delay intervals associated with surface reflections. This empirical behavior support the assumption that dominant scattering paths originate from visible surface sectors rather than from a uniformly distributed object interior, which is consistent with the surface-scattering model introduced in Sec.~\ref{sec:ideal_concept} .}

Note that three distances are needed for the localization with the principle of trilateration, while only two receives are used in the measurement campaign, so the radiation pattern of the antenna with an absorber is considered as a prior information for the estimation of the position.
Specifically, active measurements of RX1 are unavailable during time steps $[48, 100]$ and $[142, 190]$, while those of RX2 are unavailable during $[60, 100]$ and $[157,190]$.
In addition to the natural blockages caused by the movement of the human body, additional blockages were introduced. This also simulates a scenario where active measurements are preserved for energy saving. 
\begin{table} 
	\renewcommand{\baselinestretch}{1}\small\normalsize
	\setlength{\tabcolsep}{3pt} %
	\renewcommand{\arraystretch}{1.1} %
	\footnotesize
	\centering
	\caption{Average runtime per time step $n$ and average \ac{rmse} of different methods in real radio measurements.}\label{tbl:runtime_realMeas}
\begin{tabular}{lcccc}
	\toprule
	& \multicolumn{2}{c}{3 GHz} & \multicolumn{2}{c}{500 MHz} \\
	\cmidrule(lr){2-3} \cmidrule(lr){4-5}
	Method & RMSE (m) & Runtime (s) & RMSE (m) & Runtime (s) \\
	\midrule
	A-PDA       & 17.320 & 0.008 & 12.935 & 0.006 \\
	AP-PDA      & 10.050 & 0.008 & 1.775  & 0.005 \\
	A-PROP-SMP  & 1.223  & 0.132 & 1.183  & 0.131 \\
	AP-PROP-SMP & 0.134  & 0.254 & 0.209  & 0.250 \\
	\bottomrule
\end{tabular}
	\vspace{-2.5mm}
\end{table}

The performance of different methods with measured radio signals is shown in Fig.~\ref{fig:CSmeas}.
Fig.~\ref{fig:rmseMeas3G} and \ref{fig:rmseMeas500M} show the \ac{rmse} as a function of the discrete observation time $n$ with $3\,\mathrm{GHz}$ and $500\,\mathrm{MHz}$, with light gray indicating missing active measurements from one RX and dark gray indicating missing active measurements from both RXs, respectively.
Figs.~\ref{fig:cdfMeas3G} and \ref{fig:cdfMeas500M} are the \ac{ecdf} of the \ac{rmse} with $3\,\mathrm{GHz}$ and $500\,\mathrm{MHz}$, respectively.
The results demonstrate that the AP-PROP-SMP performs well with real measurement data, significantly outperforming the A-PDA, AP-PDA and A-PROP-SMP. This holds true across both considered bandwidths.
With $3\,\mathrm{GHz}$ bandwidth, the P-CRLB drops below $0.01\,\mathrm{m}$ when the \ac{los} is available\footnote{{For the real-measurement experiment, the P-CRLB* cannot be computed because it requires the true LOS amplitudes at all time step. During blockage these amplitudes are unknown, and the corresponding Fisher information cannot be evaluated.}}.
However, the \ac{rmse} achieved by AP-PROP-SMP does not fully attain the P-CRLB, possibly due to calibration errors in the measurement equipment (e.g., antennas and the channel sounder) during the campaign, or correlations among scattering components introduced during channel estimation.
With $500\,\mathrm{MHz}$ bandwidth, AP-PROP-SMP exhibts superior performance. Its \ac{rmse} closely approaches the \ac{pcrlb} and converges rapidly to it after the obstruction intervals. 
{Furthermore, Table~\ref{tbl:runtime_realMeas} compares the average runtime and \ac{rmse} per time step $n$ of all methods in this scenario at both $3\,\mathrm{GHz}$ and $500\,\mathrm{MHz}$.
A-PDA and AP-PDA achieve very low runtime, but their estimation accuracy is poor, resulting in unacceptably high \ac{rmse}. 
In contrast, the proposed AP-PROP-SMP significantly improves the estimation accuracy, reducing the average \ac{rmse} to 
$0.134\,\mathrm{m}$ at  $3\,\mathrm{GHz}$ and $0.209\,\mathrm{m}$ at $500\,\mathrm{MHz}$ even including the \ac{olos} period, while maintaining a reasonable computational cost.}
{Due to the high computational cost of Monte-Carlo integration of the EO likelihood for each particle at every time step, running AP-PROP on the real measurements becomes impractical. Therefore, only the simplified variant AP-PROP-SMP is considered.}
The evaluation in Section~\ref{sec: results_synthetic_radio} shows that the performance of AP-PDA2 is comparable to AP-PROP and AP-PROP-SMP when the additional variance is properly adapted. However, this manual tuning is highly sensitive to the object extent and signal parameters, making robust performance difficult to guarantee, especially when the spatial spread of scatter points induced by the object is large.
Therefore these additional results are omitted here for clarity.

\section{Conclusion}\label{sec:conclusion}

This paper introduces a Bayesian method for tracking a radio devices coupled with a moving agent by jointly exploiting active and passive radio measurements. The agent carrying the radio device is modeled as an \ac{eo} that scatters and attenuates signals, and a measurement model is derived that accounts for surface scattering with direction-dependent visibility. To address measurement origin uncertainty, we proposed a particle-based multiple-measurement \ac{pda} algorithm that fuses direct paths and surface-scattered components across multiple anchors. Two \ac{eo} models are presented: a physically motivated elliptical scattering model (\emph{EO model}) and a simplified geometric approximation {(\emph{simplified EO model})} for reduced complexity. Experiments with synthetic and real radio signals confirm that the proposed method improves position and extent estimation compared to point-object-based approaches, when active measurements are missing. 
The fusion of passive measurements enhance the  robustness during \ac{olos} periods, and the simplified EO model achieves a favorable trade-off between accuracy and computational cost. 

Possible directions of future work are the integration of neural-enhanced factor graphs to learn parts of the measurement and state transition model~\cite{VenLeiTerWit:JSP2023,LiaMey:TSP2024} and improve inference efficiency, as well as the development of direct radio signal-based approaches~\cite{LiaLeiMey:TSP2025} that bypass conventional measurement extraction using the raw radio signals for improved performance in complex propagation environments.


 
 \acrodef{mimo}[MIMO]{multiple input multiple output}
 \acrodef{awgn}[AWGN]{additive white Gaussian noise}
 \acrodef{bw}[BW]{bandwidth}
 \acrodef{blt}[BLT]{bluetooth}
 \acrodef{cdf}[CDF]{cumulative distribution function}
  \acrodef{ecdf}[eCDF]{empirical cumulative distribution function}
 \acrodef{crlb}[CRLB]{Cram\'er-Rao lower bound}
 \acrodef{dmc}[DMC]{dense multipath component}
 \acrodef{dut}[DUT]{device under test}
 \acrodef{eo}[EO]{extended object}
 \acrodef{eirp}[EIRP]{equivalent isotropic radiated power}
 \acrodefplural{esl}[ESLs]{electronic shelf labels} 
 \acrodef{los}[LOS]{line-of-sight}
 \acrodef{mf}[MF]{matched filter}
 \acrodef{ml}[ML]{maximum likelihood}
 \acrodef{mpc}[MPC]{multipath component}
 \acrodef{nlos}[NLOS]{non-\ac{los}}
 \acrodef{eot}[EOT]{extended object tracking}
 \acrodef{pcb}[PCB]{printed circuit board}
 \acrodef{pdf}[PDF]{probability density function}
 \acrodef{reb}[REB]{ranging error bound}
 \acrodef{rss}[RSS]{received signal strength}
 \acrodef{smc}[SMC]{specular multipath component}
 \acrodef{snr}[SNR]{signal-to-noise-ratio}
 \acrodef{sinr}[SINR]{signal-to-interference-plus-noise-ratio}
 \acrodef{tdoa}[TDOA]{time difference of arrival}
 \acrodef{tka}[TKA]{trusted keyless access}
 \acrodef{toa}[TOA]{time-of-arrival}
 \acrodef{aoa}[AOA]{angle-of-arrival}
 \acrodef{uwb}[UWB]{ultra wide band}
 \acrodef{mie}[MIE]{method of interval estimation}
 \acrodef{mc}[MC]{Monte Carlo}
 \acrodef{mse}[MSE]{mean squared error}
 \acrodef{ci}[CI]{confidence interval}
 \acrodef{cl}[CL]{confidence level}
 \acrodef{pdp}[PDP]{power delay profile}
 \acrodef{dps}[DPS]{delay power spectrum}
 \acrodef{dm}[DM]{dense multipath}
 \acrodef{nlike}[NLIKE]{normalized likelihood}
 \acrodef{zzb}[ZZB]{Ziv-Zakai bound}
 \acrodef{ut}[UT]{unscented transform}
 \acrodef{sp}[SP]{sigma point}
 \acrodef{glrt}[GLRT]{generalized likelihood ratio test}
 \acrodef{mse}[MSE]{mean squared error}
 \acrodef{rmse}[RMSE]{root mean squared error}
 \acrodef{nnlike}[NNLIKE]{normalized noise-free likelihood}
 \acrodef{stdv}[STDV]{standard deviation}
 \acrodef{rv}[RV]{random variable}
 \acrodef{bp}[BP]{belief propagation}
 \acrodef{pda}[PDA]{probabilistic data association}
 \acrodef{mp}[MP]{multipath}
 \acrodef{pmf}[PMF]{probability mass function}
 \acrodef{pdaf}[PDAF]{probabilistic data association filter}
 \acrodef{pdaai}[AIPDA]{amplitude-information \ac{pda}}
 \acrodef{olos}[OLOS]{obstructed line-of-sight}
 \acrodef{spa}[SPA]{sum-product algorithm}
 \acrodef{mmse}[MMSE]{minimum mean-square error}
 \acrodef{lhf}[LHF]{likelihood function}
 \acrodef{fa}[FA]{false alarm}
 \acrodef{ceda}[CEDA]{channel estimation and detection algorithm} 
 \acrodef{pcrlb}[P-CRLB]{posterior Cram\'er-Rao lower bound}
 \acrodef{slam}[SLAM]{simultaneous localization and mapping}
 \acrodef{mpslam}[MP-SLAM]{multipath-based SLAM}
 \acrodef{va}[VA]{virtual anchor}
 \acrodef{dnr}[DNR]{dense-to-noise ratio}
 \acrodef{aednn}[AE-DNN]{auto encoder deep neural network}   
 \acrodef{gpr}[GPR]{gaussian process regression}  
 \acrodef{ae}[AE]{auto encoder}  
 \acrodef{cir}[CIR]{channel impulse response}  
 \acrodef{fov}[FoV]{field-of-view}


\appendix\label{sec:appendix}

\subsection{Radio Signal Model}\label{sec:RSmodel}

\emph{Active Radio Signal Model:} The radio device at position $\bm{m}_{n} = \bm{p}_n+ [b_{\rho \s n} \cos(b_{\phi \s n}+\theta_n) \iist\iist b_{\rho \s n} \sin(b_{\phi \s n}+\theta_n)]\transp$ at time $n$, with $\bm{p}_n$ as center position of the \ac{eo}, $\V{b}_n = [b_{\rho\s n} \iist b_{\phi\s n}]^\text{T}$ as bias between radio device and the \ac{eo}, and $\theta_n$ as the orientation of the \ac{eo}, transmits a signal $s(t)$ that is received by the $j$th anchor, where $j \in \{1,\dots,J\}$. The complex baseband signal received at anchor $j$ is modeled as
\vspace*{-1mm}
\begin{align}
	r_{\text{A}, n}^{(j)}(t) =\alpha_{n}^{(j)}s(t-\tau_{n}^{(j)})+\sum_{i=1}^{A^{(j)}_n} \alpha_{i,n}^{(j)}s(t-\tau_{i,n}^{(j)})+ w_n^{(j)}(t)
	\label{equ:rx_active}\\[-8mm]\nn
\end{align}
where $\alpha_{n}^{(j)}$ and $\tau_{n}^{(j)}$ are the complex amplitude and delay of the LOS component from active measurements. 
The received signal of anchor $j$ also consists of $A^{(j)}_n$ \acp{mpc} originating from scatter points $\vm{q}_i \in \mathcal{S}^{(j)}(\V{X}_n,\V{x}_n)$.  
The complex amplitude and delay of the scatter component are denoted as  $\alpha_{i,n}^{(j)}$ and $\tau_{i,n}^{(j)}$.
The term $w_n^{(j)}(t)$ accounts for measurement noise modeled as \ac{awgn} with double-sided power spectral density $N_0/2$. 
The propagation delay of the LOS path and scatter paths are given respectively by $\tau_{n}^{(j)} \triangleq h^{(j)}(\bm{m}_n)/c$ and  
$\tau_{i,n}^{(j)} \triangleq h_\text{A}^{(j)}({\V{m}_n},\vm{q}_i)/c$, 
where $c$ is the speed of light, $h^{(j)}(\V{m}_n) = \|\vm{m}_n - \vm{p}_\text{a}^{(j)}\|$, and $h_\text{A}^{(j)}({\V{m}_n},\vm{q}_i) = (\|\vm{q}_i - \bm{p}_\text{a}^{(j)}\|+\|\vm{q}_i - \bm{m}_n\|)$.
The complex amplitudes of the LOS component and the $i$th scattered component observed at anchor $j$ at time $n$ are modeled as
\vspace*{-2mm}
\begin{align}
	\alpha_{n}^{(j)}
	&= \alpha_{\mathrm{A}}
	\exp{-\mathrm{i}\,2\pi f_c \tau_{n}^{(j)}}
	\frac{1}{4\pi f_c \,\tau_{n}^{(j)}}
	\label{equ:amp_active_LOS}\\[0mm]
	\alpha_{i,n}^{(j)}
	&= \alpha_{\mathrm{A}}\,\beta
	\exp{\mathrm{i}\,\phi_{i}}
	\frac{1}{4\pi f_c \,\tau_{i,n}^{(j)}}
	\label{equ:amp_active_scatter}\\[-6mm]\nn
\end{align}
where $f_c$ is the carrier frequency and $\alpha_{\mathrm{A}} = |\alpha_{\mathrm{A}}|\exp{\phi_{\mathrm{A}}}$ is a lumped complex factor capturing the transmitter/receiver chains and antenna responses with magnitude $|\alpha_{\mathrm{A}}|$ and random phase $\phi_{\mathrm{A}}\!\sim\!\mathcal{U}[0,2\pi)$. The constant coefficient $\beta$ models the amplitude reduction due to the interaction with the scatter point $\bm{q}_i$, and $\phi_{i}\!\sim\!\mathcal{U}[0,2\pi)$ denotes an additional random phase for that path. The factor $1/(4\pi f_c \tau)$ corresponds to the free-space pathloss.

\emph{Passive Radio Signal Model:} Simultaneously, a radio signal $s(t)$ is transmitted  from the $j'$th anchor with $j' \in \mathcal{A}_\text{tx}$ and received at the $j$th anchor with $j \in \{1,\dots,J\}$. The complex baseband signal received at the $j$th anchor is modeled as
\begin{equation}
	r_{\text{P}, n}^{(j,j')}(t) = \sum_{i=1}^{P_{n}^{(j,j')}}\alpha_{i,{n}}^{(j,j')}s(t-\tau_{i,{n}}^{(j,j')}) + w_{n}^{(j,j')}(t)
	\label{equ:rx_passive}
\end{equation}
where $ \alpha_{i,{n}}^{(j,j')}$ and $\tau_{i,{n}}^{(j,j')}$ are the complex amplitude and delay of  $P_{n}^{(j,j')}$ \acp{mpc} originating from scatter points $\vm{q}'_i  \in \mathcal{S}^{(j,j')}(\V{X}_n,\V{x}_n)$. Note that the direct component of the radio signal transmitted by the $j'$th anchor and received at the $j$th anchor has been subtracted from $r_{\text{P}, n}^{(j,j')}(t)$.
The propagation delay of the passive measurement is given by $\tau_{i,{n}}^{(j,j')} \triangleq h_\text{P}^{(j,j')}(\vm{q}'_i)/c$,
where $h_\text{P}^{(j,j')}(\vm{q}'_i) = (\|\vm{q}'_i - \vm{p}_\text{a}^{(j')}\| + \|\vm{q}'_i - \vm{p}_\text{a}^{(j)}\|)$.
The complex amplitude of the passive measurement between transmitting anchor $j$ and receiving anchor $j'$ is given by 
\begin{equation}
	\alpha_{i,n}^{(j,j')} = \alpha_{\text{P}}\beta \exp{\mathrm{i} \phi_i} \frac{c}{4\pi f_c \tau_{i,n}^{(j,j')}}
	\label{equ:amp_passive_scatter}
\end{equation}
where $\alpha_{\mathrm{P}} = |\alpha_{\mathrm{P}}|\exp{\phi_{\mathrm{P}}}$ is a lumped complex factor capturing the transmitter/receiver chains and antenna responses with magnitude $|\alpha_{\mathrm{P}}|$ and random phase $\phi_{\mathrm{P}}\!\sim\!\mathcal{U}[0,2\pi)$.

\subsection{Scatter Point Generation}\label{scatterPDF}
This section addresses the process of direct sampling, i.e., generating a set of scatter points
$\{\V{q}_i\}_{i=1}^{{I}}$ or $\{\V{q}'_i \}_{i=1}^{{I}}$ 
from the importance distribution  
$f(\V{q} | \V{X}_n,\V{x}_n) = \mathcal{U}(\V{q}; \mathcal{S}^{(j)}(\V{X}_n,\V{x}_n))$ or 
$f(\V{q}' | \V{X}_n,\V{x}_n) = \mathcal{U}(\V{q}'; \mathcal{S}^{(j,j')}(\V{X}_n,\V{x}_n))$ 
in order to efficiently solve \eqref{equ:measLikelihood_idealAS_appr} or \eqref{equ:measLikelihood_idealPS_appr}, respectively. 
As illustrated in Fig.~\ref{fig:Modelideal}, the regions $  \mathcal{S}^{(j)}(\V{X}_n,\V{x}_n) $ and $\mathcal{S}^{(j,j')}(\V{X}_n,\V{x}_n) $, which correspond to the blue shaded areas in Fig.~\ref{fig:ModelidealActive} and Fig.~\ref{fig:ModelidealPassive}, respectively, are sectors of an \emph{elliptic annulus}.  (i.e, an angular range of an ellipse-shaped ring), defined by inner and outer elliptic boundaries and two angular boundaries.
The sampling procedure is detailed in the following. It applies equivalently for obtaining samples $\{\V{q}_i \}_{i=1}^{{I}}$ or $\{\V{q}'_i\}_{i=1}^{{I}}$ for the active and passive measurement models.
Apart from the variables used in the previous part of this section, we also use the variables $\bm{p}_n$, {$l_n$, $w_n$, $\delta_n$}, $\varphi_{s \s n}^{(j,j')}~/~\varphi_{s \s n}^{(j)}$, and $\varphi_{e \s n}^{(j,j')}~/~\varphi_{e \s n}^{(j)}$ as defined in Section~\ref{sec:problem}
In what follows, we omit the indices for time $n$, transmitting anchor $j'$ and receiving anchor $j$ as well as the subscripts $A$ and $P$ that denote active and passive measurements for notational convenience.

We begin by sampling from a circular annulus with inner radius $ r_{\text{in}} $ and outer radius $ r_{\text{out}} $ given as 
$
r_{\text{in}}  \triangleq 1-\frac{\delta}{l}, \quad r_{\text{out}} \triangleq 1 + \frac{\delta}{l}, 
$ 
i.e., an annulus of width $\frac{\delta}{l}$ that spans symmetrically around a unit circle. 
To ensure uniform sampling over the area, we generate ${I}$ samples $i \in [ 1 ,\, \dots \, ,  {I} ]$ with $ \varphi_i \sim \mathcal{U}( \varphi_{s}, \varphi_{e}) $ and $ \rho_i  \sim \mathcal{U}(r_{\text{in}}^2, r_{\text{out}}^2) $, and define the radius as $ r_i  = \sqrt{\rho_i}  $. Samples in polar coordinates are obtained as
\begin{equation} \label{eq:deriv_sampling_1}
	\bm{q}_i^{(p)} = \sqrt{\rho_i}
	\begin{bmatrix}
		\cos(\varphi_i) \\ \sin(\varphi_i)
	\end{bmatrix} \, .
\end{equation}
This guarantees that the probability density is uniform with respect to the area in the circular domain.

\paragraph*{Proof of Area-Uniform Sampling}

If we sample $ \rho_i \sim \mathcal{U}(r_{\text{in}}^2, r_{\text{out}}^2) $, then the density of random variable $ \rho $ is
\[
f_\rho(\rho) = \left. \frac{1}{r_{\text{out}}^2 - r_{\text{in}}^2} \right|_{\rho \in [r_{\text{in}}^2, r_{\text{out}}^2]}.
\]
Using the change of variables formula, the corresponding \ac{pdf} for random variable $ r $ is
\[
f_r(r) = \left. f_\rho(r^2) \cdot \left| \frac{d}{dr}(r^2) \right| = \frac{2r}{r_{\text{out}}^2 - r_{\text{in}}^2} \right|_{r \in [r_{\text{in}}, r_{\text{out}}]}.
\]
The resulting joint \ac{pdf} in polar coordinates is given as 
\[
f_{r,\varphi}(r,\varphi) = \left. \frac{2r}{(r_{\text{out}}^2 - r_{\text{in}}^2)(\varphi_e - \varphi_s)}  \right|_{r \in [r_{\text{in}}, r_{\text{out}}],  \varphi \in [\varphi_e, \varphi_s]}.
\]
Next, we transform the \ac{pdf} from polar coordinates to Cartesian coordinates with
$x = r \cos \varphi, \quad y = r \sin \varphi$, where the Jacobian determinant is given as $J = | {\partial(x, y)}/{\partial(r, \varphi)} | = r $. We apply the change of variables formula again and obtain
\begin{align}
	f_{x,y}(x, y) &= \frac{1}{J} f_{r,\varphi}(r, \varphi)  \nn \\
	&= \left.\frac{2}{(r_{\text{out}}^2 - r_{\text{in}}^2)(\varphi_e - \varphi_s)}  \right|_{r \in [r_{\text{in}}, r_{\text{out}}],  \varphi \in [\varphi_s, \varphi_e]}. \nn
\end{align}
Hence, $ f_{x,y}(x,y) $ is constant over the annulus defined by $ r \in [r_{\text{in}}, r_{\text{out}}] $ and $ \varphi \in [\varphi_s, \varphi_e] $, which completes the proof. 

\medskip

Next, we apply an \emph{invertible linear transformation} $\bm{Q}$ to the circular annulus sector to obtain a rotated \emph{elliptic annulus sector}, given as
\begin{equation}  \label{eq:deriv_sampling_2}
	\bm{q}_i = \bm{Q} \, 	\bm{q}_i^{(p)} = 
	\begin{bmatrix}
		\cos(\theta) & -\sin(\theta) \\ \sin(\theta) & \cos(\theta) \\
	\end{bmatrix} 
	\begin{bmatrix}
		a & 0 \\ 0 & b 
	\end{bmatrix}\, 	\bm{q}_i^{(p)}
\end{equation}
with a scaling matrix that transforms the circular annulus to an elliptic annulus, and a rotation matrix with an orientation angular $\theta = \text{atan}({v_{\text{y}} }/{v_{\text{x}}})$ of the elliptic annulus. 
Linear transformations preserve the uniformity of continuous distributions over bounded regions (see e.g. \cite[Chapter 5]{papoulis2001probability}), which follows trivially by the Jacobian determinant that occurs in the change of variables being a constant for invertible linear transformations. Consequently, the resulting distribution over the elliptic sector remains uniform inside the boundaries. Combining  \eqref{eq:deriv_sampling_1} and  \eqref{eq:deriv_sampling_2} we get 
\begin{equation}  \label{eq:proposal_sampling_transform}
	\bm{q}_i= \bm{p} +
	\begin{bmatrix}
		\cos(\theta) & -\sin(\theta) \\
		\sin(\theta) & \cos(\theta)
	\end{bmatrix}
	\begin{bmatrix}
		a \sqrt{\rho_i} \cos(\varphi_i) \\
		b \sqrt{\rho_i} \sin(\varphi_i)
	\end{bmatrix}.
\end{equation}

Finally, we note that the width of the resulting elliptic annulus is not constant, since the semi-minor axis of the ellipse is effectively scaled by $w/l$. However, it is a sufficient approximation since the semi-major and semi-minor axes of the \ac{eo} model typically are of comparable length. 

In summary, the full procedure involves uniform sampling in polar coordinates over a circular annulus sector, followed by affine mapping (scaling and rotation) to obtain a uniformly sampled elliptic annulus sector.
The complete sampling procedure for obtaining a set of scatter points $\{\V{q}_i\}_{i=1}^{{I}}$ consists of
\begin{enumerate}
	\item obtaining samples $ \rho_i \sim \mathcal{U}((1-\frac{\delta}{l})^2, (1+\frac{\delta}{l})^2) $,
	\item obtaining samples $\varphi_i \sim \mathcal{U}( \varphi_{s}, \varphi_{e}) $,
	\item transforming sample pairs $(\rho_i,\varphi_i)$ with \eqref{eq:proposal_sampling_transform}.
\end{enumerate}

\subsection{Reference Method based on Point Object Assumption} \label{sec:PDA_algorithm}

%
When the radio signal bandwidth is low and, thus, the resulting pulse width is high compared to the size of the \ac{eo}, the individual scatter components in the received radio signals strongly overlap. As a result, the \ac{ceda} used for measurement extraction (see Section~\ref{sec:measurement_model}) can no longer resolve the contributions of individual scatter components \cite{Hansen2018, Grebien2024}, yielding a single measurement\footnote{Note that while there is only a single measurement caused by the \ac{eo} there still can be other measurements due to false alarms or clutter.}. Consequently, the \ac{eo} can be accurately approximated as a point object. Since we assume the radio device to be located near the surface of the \ac{eo}, the expected value of the random variable describing the extracted measurement effectively corresponds to the position of the radio device, $\bm{m}_n \equiv \bm{p}_n$. 

Based on the above assumption, the system model can be reduced to 
the kinematic state  $\bm{x}_n = [\bm{p}_n ^\text{T}\; \bm{v}_n^\text{T}]^\text{T}$, consisting of the position of the radio device $\bm{p}_n$ and its velocity $\bm{v}_n$, and the state transition of $\bm{x}_n$ still follows Section~\ref{sec:SSM}. The \ac{lhf} for active measurements that represents the contributions of the unsolvable LOS component and scatter components reduces to 
\begin{align}
	f_\text{A(pda)}(\V{z}^{(j)}_{\text{A},n,l}|\bm{x}_n) 
	= f_{\text{N}}(z_{\text{A},\text{d},n,l}^{(j)}; h^{(j)}({\V{p}}_n), \sigma_{\text{A},n,l}^{(j)\ist 2}+\sigma_{\text{r}}^2) 
	\label{equ:measLikelihood_active_point}
\end{align}
and the \ac{lhf} for passive measurements reduces to 
\begin{align}
	f_{\text{P(pda)}}(\V{z}^{(j,j')}_{\text{P},n,l}|\bm{x}_n) 
	= f_{\text{N}}(z_{\text{P}, \text{d},n,l}^{(j,j')}; h^{(j,j')}_\text{P}(\V{p}_n),  \sigma_{\text{P},n,l}^{(j,j')\ist 2} +\sigma_{\text{r}}^{2})
	\label{equ:measLikelihood_passive_point}
\end{align}
with the functions $h^{(j)}$ and  $h^{(j,j')}_\text{P}$ defined as in Section~\ref{sec:measurement_model}. 
The variance $\sigma_{\text{r}}$ captures the variance caused by the \ac{eo} and is an additional constant to be set. 
The fact that the \ac{eo} generates only a single measurement simplifies the data association problem, resulting in the association model used for conventional \ac{pda}\cite{BarShalomTCS2009}.

We obtain a state estimate of $\bm{x}_n$ using the \ac{mmse} estimate $\hat{\bm{x}}^\text{MMSE}_{n} \,\triangleq \int \rmv \bm{x}_{n} \, f(\bm{x}_{n} | \V{z}_{\text{A},1:n},\V{z}_{\text{P},1:n} )\, \mathrm{d}\bm{x}_{n} \,$, following the derivation process in \cite{Venus2024}. The resulting algorithm 
corresponds to a particle-based variant of the conventional multi-sensor \ac{pda} filter, where each set of active and passive measurements is treated as originating from an individual anchor/ anchor pair.


\section*{Acknowledgment}
The authors would like to express their sincere gratitude to Benjamin Deutschmann, Lukas D’Angelo, and Thomas Wilding for their invaluable support and assistance during the measurement campaign.


\renewcommand{\baselinestretch}{0.98}\small\normalsize 

\bibliographystyle{IEEEtran}
\bibliography{IEEEabrv, references}

\begin{thebibliography}{10}
\providecommand{\url}[1]{#1}
\csname url@samestyle\endcsname
\providecommand{\newblock}{\relax}
\providecommand{\bibinfo}[2]{#2}
\providecommand{\BIBentrySTDinterwordspacing}{\spaceskip=0pt\relax}
\providecommand{\BIBentryALTinterwordstretchfactor}{4}
\providecommand{\BIBentryALTinterwordspacing}{\spaceskip=\fontdimen2\font plus
\BIBentryALTinterwordstretchfactor\fontdimen3\font minus
  \fontdimen4\font\relax}
\providecommand{\BIBforeignlanguage}[2]{{%
\expandafter\ifx\csname l@#1\endcsname\relax
\typeout{** WARNING: IEEEtran.bst: No hyphenation pattern has been}%
\typeout{** loaded for the language `#1'. Using the pattern for}%
\typeout{** the default language instead.}%
\else
\language=\csname l@#1\endcsname
\fi
#2}}
\providecommand{\BIBdecl}{\relax}
\BIBdecl

\bibitem{Wymeersch2022}
H.~Wymeersch and G.~Seco-Granados, ``Radio localization and sensing—part i:
  Fundamentals,'' vol.~26, no.~12, pp. 2816--2820, 2022.

\bibitem{WitrisalSPM2016Copy}
K.~Witrisal, P.~Meissner \emph{et~al.}, ``High-accuracy localization for
  assisted living: {5G} systems will turn multipath channels from foe to
  friend,'' \emph{{IEEE} Signal Process. Mag.}, vol.~33, no.~2, pp. 59--70,
  Mar. 2016.

\bibitem{Gedschold2023}
J.~Gedschold, S.~Semper, R.~S. Thom{\"a}, M.~Döbereiner, and G.~D. Galdo,
  ``Dynamic delay-dispersive {UWB}-radar targets: Modeling and estimation,''
  \emph{{IEEE} Trans. Antennas Propag.}, vol.~71, no.~8, pp. 6814--6829, 2023.

\bibitem{Zetik2007}
R.~Zetik, J.~Sachs, and R.~S. Thoma, ``{UWB} short-range radar sensing - the
  architecture of a baseband, pseudo-noise {UWB} radar sensor,'' \emph{{IEEE}
  Instrum. Meas. Mag.}, vol.~10, no.~2, pp. 39--45, 2007.

\bibitem{PonteMueller2023}
F.~de~Ponte~Müller, M.~Schmidhammer, and S.~Sand, ``Radio-based sensing in
  vehicular environments: Robust localization and tracking of {VRU}s,'' in
  \emph{Proc. IEEE 97th Veh. Technol. Conf. (VTC2023-Spring)}, 2023, pp. 1--6.

\bibitem{ZhaStaJosWanGenDamWymHoeTAES2020}
S.~Zhang, E.~Staudinger, T.~Jost, W.~Wang, C.~Gentner, A.~Dammann,
  H.~Wymeersch, and P.~A. Hoeher, ``Distributed direct localization suitable
  for dense networks,'' \emph{{IEEE} Trans. Aerosp. Electron. Syst.}, vol.~56,
  no.~2, pp. 1209--1227, July 2020.

\bibitem{Mendrzik2019}
R.~Mendrzik, H.~Wymeersch, G.~Bauch, and Z.~Abu-Shaban, ``{H}arnessing {NLOS}
  {C}omponents for {P}osition and {O}rientation {E}stimation in {5G}
  {M}illimeter {W}ave {MIMO},'' \emph{{IEEE} Trans. Wireless Commun.}, vol.~18,
  no.~1, pp. 93--107, 2019.

\bibitem{LeiVenTeaMey:TSP2023}
E.~{Leitinger}, A.~{Venus}, B.~{Teague}, and F.~{Meyer}, ``Data fusion for
  multipath-based {SLAM}: {Combining} information from multiple propagation
  paths,'' \emph{{IEEE} Trans. Signal Process.}, vol.~71, pp. 4011--4028, Sep.
  2023.

\bibitem{Ambroziak2016}
S.~J. Ambroziak, L.~M. Correia, R.~J. Katulski, M.~Mackowiak, C.~Oliveira,
  J.~Sadowski, and K.~Turbic, ``An off-body channel model for body area
  networks in indoor environments,'' \emph{{IEEE} Trans. Antennas Propag.},
  vol.~64, no.~9, pp. 4022--4035, 2016.

\bibitem{GraBau:JAIF2017}
K.~Granstr{\"{o}}m and M.~Baum, ``Extended object tracking: {Introduction},
  overview and applications,'' \emph{J. Adv. Inf. Fusion}, vol.~12, no.~2, pp.
  139--174, Dec. 2017.

\bibitem{Baum2014}
M.~Baum and U.~D. Hanebeck, ``Extended object tracking with random hypersurface
  models,'' \emph{{IEEE} Trans. Aerosp. Electron. Syst.}, vol.~50, no.~1, pp.
  149--159, 2014.

\bibitem{Hirscher2016}
T.~Hirscher, A.~Scheel, S.~Reuter, and K.~Dietmayer, ``Multiple extended object
  tracking using {Gaussian} processes,'' in \emph{Proc. Fusion-2016},
  Heidelberg, Germany, Jul. 2016, pp. 868--875.

\bibitem{Kumru2021}
M.~Kumru and E.~{\"O}zkan, ``Three-dimensional extended object tracking and
  shape learning using {Gaussian} processes,'' \emph{{IEEE} Trans. Aerosp.
  Electron. Syst.}, vol.~57, no.~5, pp. 2795--2814, 2021.

\bibitem{Koch2008}
J.~W. Koch, ``Bayesian approach to extended object and cluster tracking using
  random matrices,'' \emph{{IEEE} Trans. Aerosp. Electron. Syst.}, vol.~44,
  no.~3, pp. 1042--1059, 2008.

\bibitem{Feldmann2011}
M.~Feldmann, D.~Fr{\"a}nken, and W.~Koch, ``Tracking of extended objects and
  group targets using random matrices,'' \emph{{IEEE} Trans. Signal Process.},
  vol.~59, no.~4, pp. 1409--1420, 2011.

\bibitem{Schuster2015}
M.~Schuster, J.~Reuter, and G.~Wanielik, ``Probabilistic data association for
  tracking extended targets under clutter using random matrices,'' in
  \emph{Proc. Fusion-2015}, Washington, DC, USA, Jul. 2015, pp. 961--968.

\bibitem{Zhang2021}
L.~Zhang and J.~Lan, ``Tracking of extended object using random matrix with
  non-uniformly distributed measurements,'' \emph{{IEEE} Trans. Signal
  Process.}, vol.~69, pp. 3812--3825, 2021.

\bibitem{Hoher2022}
P.~Hoher, S.~Wirtensohn, T.~Baur, J.~Reuter, F.~Govaers, and W.~Koch,
  ``Extended target tracking with a lidar sensor using random matrices and a
  virtual measurement model,'' \emph{{IEEE} Trans. Signal Process.}, vol.~70,
  pp. 228--239, Dec. 2022.

\bibitem{Granstroem2011}
K.~Granstr{\"o}m, C.~Lundquist, and U.~Orguner, ``Tracking rectangular and
  elliptical extended targets using laser measurements,'' in \emph{Proc.
  Fusion-2011}, Chicago, IL, USA, Jul. 2011, pp. 1--8.

\bibitem{Granstroem2014}
K.~Granstr{\"o}m, S.~Reuter, D.~Meissner, and A.~Scheel, ``A multiple model
  {PHD} approach to tracking of cars under an assumed rectangular shape,'' in
  \emph{Proc. Fusion-2014}, Salamanca, Spain, Jul. 2014, pp. 1--8.

\bibitem{BarShalomTCS2009}
Y.~Bar-Shalom, F.~Daum, and J.~Huang, ``The probabilistic data association
  filter,'' \emph{{IEEE} Control Syst. Mag.}, vol.~29, no.~6, pp. 82--100, Dec
  2009.

\bibitem{JeoTugTAES2005}
S.~Jeong and J.~Tugnait, ``Multisensor tracking of a maneuvering target in
  clutter using {IMMPDA} filtering with simultaneous measurement update,''
  \emph{{IEEE} Trans. Aerosp. Electron. Syst.}, vol.~41, no.~3, pp. 1122--1131,
  Nov. 2005.

\bibitem{Venus2021}
A.~Venus, E.~Leitinger, S.~Tertinek, and K.~Witrisal, ``A message passing based
  adaptive {PDA} algorithm for robust radio-based localization and tracking,''
  in \emph{2021 Proc. IEEE RadarConf-21}, 2021, pp. 1--6.

\bibitem{MeyerTSP2021}
F.~Meyer and J.~L. Williams, ``Scalable detection and tracking of geometric
  extended objects,'' \emph{{IEEE} Trans. Signal Process.}, vol.~69, pp.
  6283--6298, 2021.

\bibitem{Wielandner2024}
L.~Wielandner, A.~Venus, T.~Wilding, K.~Witrisal, and E.~Leitinger, ``{MIMO}
  multipath-based {SLAM} for non-ideal reflective surfaces,'' in \emph{Proc.
  Fusion-2024}, Venice, Italy, Jul. 2024.

\bibitem{WildingPIMRC2020}
T.~{Wilding}, E.~{Leitinger}, U.~{Muehlmann}, and K.~{Witrisal}, ``Modeling
  human body influence in {UWB} channels,'' in \emph{Proc. IEEE 31st Annu. Int.
  Symp. Pers., Indoor Mobile Radio Commun. (PIMRC)}, London, UK, Aug. 2020, pp.
  1--6.

\bibitem{Zhu2024}
H.~Zhu, A.~Venus, E.~Leitinger, S.~Tertinek, and K.~Witrisal, ``Fusion of
  active and passive measurements for robust and scalable positioning,'' in
  \emph{Proc. IEEE 4th Int. Symp. Joint Commun. \& Sens. (JC\&S)}, Leuven,
  Belgium, Mar. 2024, pp. 1--6.

\bibitem{Zhu2025}
\BIBentryALTinterwordspacing
H.~Zhu, A.~Venus, E.~Leitinger, and K.~Witrisal, ``Multi-sensor fusion of
  active and passive measurements for extended object tracking,'' 2025.
  [Online]. Available: \url{https://arxiv.org/abs/2504.18301}
\BIBentrySTDinterwordspacing

\bibitem{Kay1998}
S.~Kay, \emph{Fundamentals of Statistical Signal Processing: Detection
  Theory}.\hskip 1em plus 0.5em minus 0.4em\relax Upper Saddle River, NJ, USA:
  Prentice Hall, 1998.

\bibitem{BarShalom2002EstimationTracking}
Y.~Bar-Shalom, T.~Kirubarajan, and X.-R. Li, \emph{Estimation with Applications
  to Tracking and Navigation}.\hskip 1em plus 0.5em minus 0.4em\relax New York,
  NY, USA: John Wiley \& Sons, Inc., 2002.

\bibitem{Hansen2018}
T.~L. Hansen, B.~H. Fleury, and B.~D. Rao, ``Superfast line spectral
  estimation,'' \emph{{IEEE} Trans. Signal Process.}, vol.~PP, no.~99, pp.
  1--1, Feb. 2018.

\bibitem{Grebien2024}
S.~Grebien, E.~Leitinger, K.~Witrisal, and B.~H. Fleury, ``Super-resolution
  estimation of {UWB} channels including the dense component -- {An
  SBL}-inspired approach,'' \emph{{IEEE} Trans. Wireless Commun.}, vol.~23,
  no.~8, pp. 10\,301--10\,318, Feb. 2024.

\bibitem{WitrisalJWCOML2016}
K.~Witrisal, E.~Leitinger, S.~Hinteregger, and P.~Meissner, ``Bandwidth scaling
  and diversity gain for ranging and positioning in dense multipath channels,''
  \emph{{IEEE} Wireless Commun. Lett.}, vol.~5, no.~4, pp. 396--399, Aug. 2016.

\bibitem{LeitingerJSAC2015}
E.~Leitinger, P.~Meissner, C.~Rudisser, G.~Dumphart, and K.~Witrisal,
  ``Evaluation of position-related information in multipath components for
  indoor positioning,'' \emph{{IEEE} J. Sel. Areas Commun.}, vol.~33, no.~11,
  pp. 2313--2328, Nov. 2015.

\bibitem{DouWan:SPM2005}
D.~A. and W.~X., ``Monte carlo methods for signal processing: a review in the
  statistical signal processing context,'' \emph{{IEEE} Signal Process. Mag.},
  vol.~22, no.~6, pp. 152--170, Nov. 2005.

\bibitem{Ristic2003}
B.~Ristic, S.~Arulampalam, and N.~Gordon, \emph{Beyond the {K}alman filter:
  {P}article filters for tracking applications}.\hskip 1em plus 0.5em minus
  0.4em\relax Artech House, 2003.

\bibitem{Kay1993}
S.~Kay, \emph{Fundamentals of Statistical Signal Processing: Estimation
  Theory}.\hskip 1em plus 0.5em minus 0.4em\relax Upper Saddle River, NJ, USA:
  Prentice Hall, 1993.

\bibitem{KschischangTIT2001}
F.~Kschischang, B.~Frey, and H.-A. Loeliger, ``Factor graphs and the
  sum-product algorithm,'' \emph{{IEEE} Trans. Inf. Theory}, vol.~47, no.~2,
  pp. 498--519, Feb. 2001.

\bibitem{MeyerJSIPN2016}
F.~Meyer, O.~Hlinka, H.~Wymeersch, E.~Riegler, and F.~Hlawatsch, ``Distributed
  localization and tracking of mobile networks including noncooperative
  objects,'' \emph{{IEEE} Trans. Signal Inf. Process. Netw.}, vol.~2, no.~1,
  pp. 57--71, 2016.

\bibitem{ArulampalamTSP2002}
M.~S. Arulampalam, S.~Maskell, N.~Gordon, and T.~Clapp, ``A tutorial on
  particle filters for online nonlinear/non-{Gaussian} {Bayesian} tracking,''
  \emph{{IEEE} Trans. Signal Process.}, vol.~50, no.~2, pp. 174--188, Feb.
  2002.

\bibitem{Tichavsky1998}
P.~Tichavsky, C.~Muravchik, and A.~Nehorai, ``Posterior {Cramer-Rao} bounds for
  discrete-time nonlinear filtering,'' \emph{{IEEE} Trans. Signal Process.},
  vol.~46, no.~5, pp. 1386--1396, May 1998.

\bibitem{Venus2024}
A.~{Venus}, E.~{Leitinger}, S.~{Tertinek}, F.~{Meyer}, and K.~{Witrisal},
  ``Graph-based simultaneous localization and bias tracking,'' \emph{{IEEE}
  Trans. Wireless Commun.}, vol.~23, no.~10, pp. 13\,141--13\,158, May 2024.

\bibitem{LiTWC2022}
X.~Li, E.~Leitinger, A.~Venus, and F.~Tufvesson, ``Sequential detection and
  estimation of multipath channel parameters using belief propagation,''
  \emph{{IEEE} Trans. Wireless Commun.}, pp. 1--1, 2022.

\bibitem{Costa2009}
J.~R. Costa, C.~R. Medeiros, and C.~A. Fernandes, ``Performance of a crossed
  exponentially tapered slot antenna for {UWB} systems,'' \emph{{IEEE} Trans.
  Antennas Propag.}, vol.~57, no.~5, pp. 1345--1352, 2009.

\bibitem{VenLeiTerWit:JSP2023}
A.~Venus, E.~Leitinger, S.~Tertinek, and K.~Witrisal, ``A neural-enhanced
  factor graph-based algorithm for robust positioning in obstructed {LOS}
  situations,'' \emph{IEEE Open J. of Signal Process.}, vol.~5, pp. 29--38,
  Nov. 2023.

\bibitem{LiaMey:TSP2024}
M.~Liang and F.~Meyer, ``Neural enhanced belief propagation for multiobject
  tracking,'' \emph{{IEEE} Trans. Signal Process.}, vol.~72, pp. 15--30, Sept.
  2023.

\bibitem{LiaLeiMey:TSP2025}
M.~Liang, E.~Leitinger, and F.~Meyer, ``Direct multipath-based {SLAM},''
  \emph{{IEEE} Trans. Signal Process.}, vol.~73, pp. 2336--2352, 2025.

\bibitem{papoulis2001probability}
A.~Papoulis and S.~Pillai, \emph{Probability, Random Variables, and Stochastic
  Processes}.\hskip 1em plus 0.5em minus 0.4em\relax McGraw-Hill, 2001.

\end{thebibliography}

\end{document}